\newif\ifanonymous
\newif\ifdraft

\anonymousfalse
\draftfalse
\InputIfFileExists{localflags}{}

\ifdraft
\fi

\documentclass[sigconf,authorversion,nonacm,natbib=false]{acmart}

\usepackage{versions}
\includeversion{full}
\excludeversion{conf}

\usepackage[
    backend=biber,
    style=ieee
]{biblatex}

\usepackage{colortbl}
\usepackage{array}
\usepackage{pifont}
\usepackage{enumitem}
\usepackage{graphicx}
\usepackage{amsmath}
\usepackage{xr} 
\usepackage{hyperref}
\usepackage{cleveref}
\crefname{lstlisting}{listing}{listings}
\Crefname{lstlisting}{Lst}{Lst}
\usepackage[dvipsnames]{xcolor}
\usepackage{tikz}
\usepackage{multirow}
\usepackage{booktabs}
\usepackage{float}
\usepackage{textcomp}

\usepackage{sourcecodepro} 

\usetikzlibrary{positioning, fit}

\newtheorem{lemma}{Lemma}

\usepackage{listings}

\usepackage{macros-basic}
\usepackage{macros-dy-theory}
\usepackage{macros-msr}
\usepackage{macros-listings-tamarin}
\input{macros-local.sty}
\input{macros-fulltricks.sty}

\DeclareTextCommandDefault{\copyright}{\textcopyright} 

\begin{document}

\lstMakeShortInline|

\newcommand{\cmark}{\textcolor{green!70!black}{\ding{51}}} 
\newcommand{\xmark}{\textcolor{red}{\ding{55}}\hspace{2pt}} 
\newcommand{\omark}{\textcolor{orange}{\ding{108}}}        

\newcommand{\changed}[1]{\textcolor{red}{#1}}


\title{Accountability in Certificate Transparency and Variants}

\ifanonymous
    \author{}
\else
\author{Timo Treitz}
\orcid{0009-0009-3531-0061}
\affiliation{
    \institution{Saarland University}
    \city{Saarbr\"ucken}
    \country{Germany}
}
\email{titr00001@stud.uni-saarland.de}

\author{Robert K\"unnemann}
\orcid{0000-0003-0822-9283}
\affiliation{
    \institution{CISPA Helmholtz Center for Information Security}
    \city{Saarbr\"ucken}
    \country{Germany}
}
\email{robert.kuennemann@cispa.de}

\fi

\begin{abstract} \label{sec:abstract}
Certificate Transparency (CT) aims to reduce the trust required in Certificate Authorities (CAs) within the TLS certificate ecosystem. It is supported by all major browsers.
The protocol obliges all CAs to record  the certificates they issue in
a public log, which itself is monitored for compliance and
consistency by third parties.
Given this complex set of checks between the four roles—CA, loggers, monitor but also the end user's client—it is very hard to  provide a precise account of how CT eliminates trust assumptions in exchange for complex infrastructure. Analyses both in the Dolev-Yao paradigm and the computational paradigm only
regard a very simplified model and feature definitions adapted
specifically to CAs, essentially capturing design features rather than
the target property.

The present paper posits accountability as the main goal of CT and
presents a thorough analysis in the Dolev-Yao model.
We start with the vanilla PKI and, step by step, move to CT, finally
analyzing proposed extensions for SCT Auditing and Gossiping.
We show that plain CT relies on an honest log, but provides accountability
under this assumption. Furthermore, we show that the SCT Auditing extension can
eliminate this assumption, while the Gossiping extension cannot.
\end{abstract}

\maketitle

\section{Introduction}
\label{sec:introduction}

The web relies heavily on its Public Key Infrastructure (PKI) to ensure secure communication and authenticitcy. 
The web PKI utilizes certificates issued by Certificate Authorities (CAs) to establish trusted connections between clients and websites, as specified in the TLS protocol \cite{TLS-RFC8446}. 
Historically, CAs act as trusted third parties that assert (most importantly) the hostname-key mapping presented in the certificate. CAs are expected to follow rigorous guidelines, the CA/Browser Forum Baseline Requirements \cite{BaselineRequirementsIssuance2025} to prevent wrongly issued certificates. 
There have been multiple instances where CAs violated these requirements, which at worst allowed attackers to perform man-in-the-middle attacks \cite{CAEntrustIssues, amannMissionAccomplishedHTTPS2017,mozillaRevokingTrustTwo2013,obrienChromesPlanDistrust2017,sleeviSustainingDigitalCertificate2015}. Closely monitoring the actions of CAs has been a long-standing challenge with hundreds of CAs and tens of millions of daily issued certificates \cite{MerkleTown2025}.
To address this issue, the Certificate Transparency (CT) protocol was introduced \cite{laurieCertificateTransparency2014,laurieCertificateTransparency2013}, which enjoys ever-growing adoption in more and more browsers and most recently found its way into the Android ecosystem \cite{UserAgentsCertificaten.d.}. 
CT requires all issued certificates to be published in append-only logs, that allow anyone to monitor and audit the certificates issued by CAs. This effectively distributes the trust assumption from a single CA to the CA and involved logs. CT does not prevent malicious behavior, but it aims at making it visible early on. The steps after detection of such an attack are outside the scope of CT, but CT logs should provide the necessary information to identify the attack and attribute it to the responsible CA. A successful example of this was the case against Symantec, where multiple rogue certificates were found in CT logs \cite{obrienChromesPlanDistrust2017, sleeviSustainingDigitalCertificate2015}.
We posit accountability, the ability to identify and correctly attribute misbehavior to the responsible parties, as the key property of CT.

We apply automated analysis to CT using Tamarin in the Dolev-Yao model. Tamarin has built-in support to prove accountability: we define a set of \emph{case tests} that aim to identify parties violating the security property. This mechanism is part of the protocol, although often underspecified. It is implemented through the parties in the protocol, without an outside view of every action. We verify, formally and automatically, that this identification contains all parties it should and no party it should not.
Our model of CT is the most detailed to date, and the
first formal model able to describe how to keep logs or monitors accountable. Our results are the first to consider corruption scenarios
involving all three roles: CA, log, and monitor. Compared to results outside formal methods, we are not only the first to be able to hold logs or monitors accountable, but the first to guarantee that (a) all responsible parties are held accountable (not just one) and (b) no innocent party is blamed. 
Our goal is to show that if authenticity is violated, CT provides the necessary information to identify the misbehaving parties causing the violation.
First, we find attacks refuting this claim. 
Second, we formulate assumptions that prove it but are unrealistic. Third, we show how additional infrastructure (in
CT, or, later, in extension of CT) may ensure these assumptions.
We perform this detailed analysis for basic CT, for SCT auditing
(as implemented in the Chrome browser) and STH gossiping (an
academic proposal).
%

\begin{full}%
\subsubsection*{Organisation}
In \Cref{sec:background}, we start with necessary background, including the CT standard and its proposed extensions. \Cref{sec:methodology} gives a brief overview of modeling using Tamarin and further details on the accountability notion we use in this work.
We then start incrementally introducing our model and consider the PKI setting first in \Cref{sec:pki}, where we show that authenticity can be violated and accountable authenticity achieved only given an external validator, which is not applicable in practice.
\Cref{sec:CT} then adds Certificate Transparency to the PKI model. We show that authenticity is still violated and that accountable authenticity requires transparency, which can be violated too.
In \Cref{sec:ctwaudit,sec:ctwgossip} we consider two additional proposals, SCT auditing and gossiping, and attempt to prove the same properties. We find that SCT auditing requires the fewest additional assumptions to achieve accountability.
\Cref{sec:evaluation} recapitulates our findings and presents details on the proofs for every considered property and model.
\Cref{sec:related-work} summarizes other works that consider CT and accountability.
\end{full}

\section{Certificate Transparency Protocol}\label{sec:background}
\label{sec:browser_impls}

\begin{figure*}
\centering
\resizebox{0.9\linewidth}{!}{
\begin{tikzpicture}[thick,scale=1, every node/.style={scale=1.2}]
  \pgfmathsetmacro{\xshift}{6} 

  \path[use as bounding box] (1,1) rectangle (27,10);

  \draw[thick, dash pattern=on 8pt off 3pt] (\xshift,1.5) rectangle (14 + \xshift,10);
  \node[anchor=center] at (1.75 + \xshift,9.75)     {\textbf{WebPKI with CT}};
  \node[anchor=center] at (2.5 + \xshift,8.5) {\includegraphics[width=2cm]{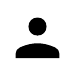}};
  \node[anchor=center] at (2.5 + \xshift,7.3)     {Domain Owner};

  \node[anchor=center] at (11.5 + \xshift,8.5) {\includegraphics[width=2cm]{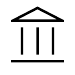}};
  \node[anchor=center] at (11.5 + \xshift,7.3)     {CA};

  \node[anchor=center] at (6.5 + \xshift,5.5) {\includegraphics[width=2cm]{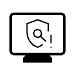}};
  \node[anchor=center] at (6.5 + \xshift,4.4)     {Monitor};

  \node[anchor=center] at (2.5 + \xshift,3) {\includegraphics[width=2cm]{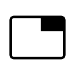}};
  \node[anchor=center] at (2.5 + \xshift,1.9)     {Browser};

  \node[anchor=center] at (11.5 + \xshift,3) {\includegraphics[width=1.7cm]{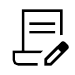}};
  \node[anchor=center] at (11.5 + \xshift,1.9)     {Log operator};

  \node[anchor=center] at (10.0 + \xshift,3.1) {\includegraphics[width=1.2cm]{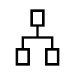}};
  \node[anchor=center, align=center] at (10.0 + \xshift,2.3)     {\scriptsize Merkle};
  \node[anchor=center, align=center] at (10.0 + \xshift,2.0)     {\scriptsize Tree};

  \draw[->, thick] (3.5 + \xshift,8.6) -- (10.5 + \xshift,8.6); 
  \node[anchor=center] at (7 + \xshift,9.0)     {\small requests certificate};
  \node[anchor=center] at (5 + \xshift,9.0) {\includegraphics[width=0.55cm]{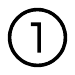}};
  \draw[<-, thick] (3.5 + \xshift,8.3) -- (10.5 + \xshift,8.3); 
  \node[anchor=center] at (7 + \xshift,7.9)     {\small issues finished cert chain with SCTs};
  \node[anchor=center] at (4.2 + \xshift,7.9) {\includegraphics[width=0.55cm]{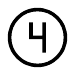}};

  \draw[->, thick] (11.7 + \xshift,7.0) -- (11.7 + \xshift,4); 
  \node[anchor=center, align=center] at (13.0 + \xshift,5.5)     {\small issues};
  \node[anchor=center, align=center] at (13.0 + \xshift,5.1)     {\small PreCertificate};
  \node[anchor=center] at (13.0 + \xshift,6) {\includegraphics[width=0.55cm]{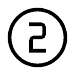}};

  \draw[<-, thick] (11.3 + \xshift,7.0) -- (11.3 + \xshift,4); 
  \node[anchor=center, align=center] at (10.0 + \xshift,5.5)     {\small responds with};
  \node[anchor=center, align=center] at (10.0 + \xshift,5.1)     {\small SCT};
  \node[anchor=center] at (10.0 + \xshift,6) {\includegraphics[width=0.55cm]{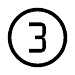}};

  \draw[->, thick] (2.5 + \xshift,7.0) -- (2.5 + \xshift,4); 
  \node[anchor=center, align=center] at (1.3 + \xshift,5.5)     {\small presents cert};
  \node[anchor=center, align=center] at (1.3 + \xshift,5.1)     {\small chain with SCTs};
  \node[anchor=center] at (1.3 + \xshift,6) {\includegraphics[width=0.55cm]{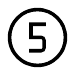}};

  \draw[<-, thick] (10.3 + \xshift,3.5) -- (11.0 + \xshift,3.5); 

  \draw[->, thick] (3.5 + \xshift,3) -- (9.5 + \xshift,3); 
  \node[anchor=center] at (6.5 + \xshift,2.3)     {\small may verify inclusion};
  \node[anchor=center] at (6.5 + \xshift,2.9) {\includegraphics[width=1cm]{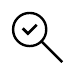}};

  \draw[->, thick] (7.5 + \xshift,4.85) -- (9.5 + \xshift,3.6); 
  \node[anchor=center] at (8.2 + \xshift,3.8)     {\small inspects};
  \node[anchor=center] at (8.2 + \xshift,3.5)     {\small logs};
  \node[anchor=center] at (8.2 + \xshift,4.3) {\includegraphics[width=0.9cm]{imgs/mag.pdf}};

  \draw[<-, thick] (4.0 + \xshift,7) -- (5.5 + \xshift,6.05); 
  \node[anchor=center] at (4.3 + \xshift,6.3)     {\small notifies};

  \draw[thick, dash pattern=on 8pt off 3pt] (1,1.5) rectangle (5.5,10);
  \draw[fill=white, color=white] (5.0,2.23) rectangle (6.5,3.32);
  \node[anchor=center] at (2.5,9.75)     {\textbf{SCT Auditing}};

  \node[anchor=center] at (3, 3) {\includegraphics[width=2cm]{imgs/Monitor.pdf}};
  \node[anchor=center] at (3, 1.9)     {Monitor (Google)};

  \node[anchor=center] at (3,8.5) {\includegraphics[width=1.7cm]{imgs/logger.pdf}};
  \node[anchor=center] at (3,7.3)     {Log operator};

  \draw[->, thick] (3,4) -- (3,7); 
  \draw[fill=white, color=white] (2.0,6.0) rectangle (4,5);
  \node[anchor=center] at (3,6.0) {\includegraphics[width=1cm]{imgs/mag.pdf}};
  \node[anchor=center, align=center] at (3,5.25)     {\small verifies SCT's claim};
  
  \draw[->, thick] (7.3,3) -- (4.3,3); 
  \node[anchor=center, align=center] at (5.8,2.75)     {\small shares full cert chain};
  
  \draw[thick, dash pattern=on 8pt off 3pt] (20.5,1.5) rectangle (26.5,10);
  
  \node[anchor=center] at (23.5,9.75)     {\textbf{CA Certification and Attestation}};
  
  \node[anchor=center] at (22.,8.5) {\includegraphics[width=2cm]{imgs/CA.pdf}};
  \node[anchor=center] at (22.,7.3)     {Root CA};

  \node[anchor=center] at (22.,3) {\includegraphics[width=1.7cm]{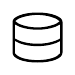}};
  \node[anchor=center] at (22.,1.9)     {CCADB};

  \node[anchor=center] at (25.0,5.5) {\includegraphics[width=1.7cm]{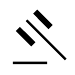}};
  \node[anchor=center] at (25.0,4.4)     {Legal Auditor};

  \draw[->, thick] (22.,7) -- (22.,4); 

  \draw[fill=white, color=white] (21., 7.75) rectangle (18.5,8.75);
  \draw[->, thick, dotted] (21.,8.5) -- (18.5,8.5); 
  \node[anchor=center] at (19.75,8.25)     {\small gets certified};
  \node[anchor=center] at (19.75,7.95)     {\tiny (out-of-band)};

  \draw[fill=white, color=white] (22.0,6.0) rectangle (24,5.2);
  \node[anchor=center, align=center] at (22,5.75)     {\small submits};
  \node[anchor=center, align=center] at (22,5.45)     {\small attestation data};

  \draw[<->, thick] (23.,7.5) -- (24.5,6.0); 
  \node[anchor=center, align=center] at (25.,7.2)     {\small attestation process};
  \node[anchor=center, align=center] at (25.,6.9)     {\small (yearly)};



\end{tikzpicture}
}
\caption{Different roles and usual information flow in the CT protocol with pre-certificates. The left box highlights the additional information flow for SCT auditing. The right box shows out-of-band steps of the WebPKI as outlined in the Baseline Requirements.}
\label{fig:CT_fullproto}
\end{figure*}

\begin{table*}
\centering
\small
\begin{tabular}{@{}lllllll@{}}
\toprule
~& Chrome~\cite{ChromeCertificateTransparency,OptinSCTAuditing}
 & Edge
 & Firefox \cite{SecurityEngineeringCertificateTransparency}\cite{FirefoxReleaseNotes}
 & Safari~\cite{ApplesCertificateTransparency2025}
 & Brave$^3$~\cite{TLSPolicy2024}  \\
\midrule
    CT supported / enforced & \cmark        & \cmark $^2$       & \cmark       & \cmark        & \cmark        \\

client-side inclusion check   & \xmark        & ---               & \xmark        & \xmark     & ---  \\

SCT Audit~\cite{OptinSCTAuditing}   & \omark$^1$       & ---        & \xmark              & \xmark   & --- \\
\# of required SCTs & 2~$^4$ & --- & 2~$^4$ & 2~$^4$ & 2~$^{3,4}$  \\
\# of trusted loggers & 7 & --- & 7 & 8 & 7--8~$^3$ \\
\bottomrule
\multicolumn{7}{l}{
    \footnotesize
$^1$~for random set of certificates
$^2$ no public policy, hence tested 
$^3$ follows policy of Apple or Chromium, depending on the platform}
\\
\multicolumn{7}{l}{
    \footnotesize
$^4$ 3 if validity > 180 days but only 2 from distinct log operators. 
}
\\
%



\end{tabular}
\caption{CT features per Browsers according to their documentation and our testing. \omark~indicates that the feature is supported with some constraints. --- shows that there was no information available.}
\label{tab:browser_CT_impls}
\end{table*}

Root CAs form the root of trust in the plain PKI setting, which are implicitly trusted by all parties.
They sign certificates for intermediate CAs that are then transitively trusted as well and sign certificates for end entities. 
In order to verify a certificate of an end entity, the entire chain of certificates up to a root CA certificate is verified (chain-of-trust). 
There are hundreds of root CAs, each being the single point of failure. The requirements for a root CA are very strict. 
Following the baseline requirements, they undergo yearly legal audits and publish all their signed intermediate CAs to the public CCADB database \cite{BaselineRequirementsIssuance2025,CommonCADatabase}.
In order to distribute trust and make the issuance of certificates transparent, Certificate Transparency (CT) introduces two new roles to the PKI ecosystem: logs and monitors \cite{laurieCertificateTransparency2013}.

\emph{Logs} are responsible for maintaining a binary Merkle tree of all submitted certificates. The data structure can be efficiently checked for append-only and inclusion. Append-only means that a log between two commits to its Merkle tree, i.e., two \emph{Signed Tree Heads} (STHs), can only ever add new certificates, but never remove or modify existing ones. 
The proof is calculated by presenting missing hashes of new entries that were added after the first STH. The old STH combined with these missing hashes should then yield the new STH.
Inclusion means that a certificate is included in the log, which can be proven by providing the hashes of the siblings of the Merkle tree nodes on the path from the leaf to the root.
Everyone can submit certificates to a log, but typically the signing CA does so by submitting a \emph{pre-certificate}, a certificate whose sole purpose is to be included in a log but not yet accepted by other parties. Besides being a pre-certificate, it contains all the fields of the final certificate that will be issued later.
On submission, the log replies with a \emph{Signed Certificate Timestamp} (SCT). It contains the timestamp of submission and the log operator's identity. The SCT is signed where the signature is calculated over the SCT's content and the contents of the certificate the SCT is issued for. 
This allows one to verify the authenticity of an SCT and attribute it to a certificate.
An SCT is a promise by the log to include the certificate together with its entire chain of CA certificates required to verify it in its Merkle tree within the \emph{Maximum Merge Delay} (MMD), which is typically up to 24 hours. There are different ways to serve the SCTs to clients, but most commonly in practice, the CA includes the SCT in the final certificate \cite{blagovStateCertificateTransparency2020,MerkleTown2025}.

\emph{Monitors} are responsible for continuous checking the logs in order to find rogue certificates. They fetch all certificates contained in a Merkle tree while also verifying the append-only property. Everyone can run a monitor or task a third party to do so. Third-party monitors scan logs for certificates that match the domain name of their customers to alert them \cite{ctprojectMonitorsCertificateTransparencyn.d.}. The decision as to whether a certificate is rogue depends on the customers' knowledge. We present all roles in the PKI system with CT in \Cref{fig:CT_fullproto}.

\subsubsection*{Known attacks and their detection}

CT does not prevent malicious CAs from issuing rogue certificates, but it makes the issuance of such certificates transparent and thus detectable as long as the involved logs do not collude with the CA.

CT introduces new attack vectors for malicious logs, such as the partition attack, where a log can present different views of the Merkle tree to different parties \cite{10.1007/978-3-319-76481-8_13}. Each party can verify the append-only property, but this only verifies consistency with party's previous views, not consistency with other parties's views.
Equivocation is only detectable if different parties compare their views of the log, a tactic known as \emph{Gossiping} \cite{nordbergGossipingCT2018}.
The omission attack violates transparency by issuing an SCT for a certificate without ever including it in the log. This attack is detectable by obtaining a failed inclusion proof together with an SCT that promised the opposite. 
Monitors that look for omitted rogue certificates will fail to detect the attack as long as they are not aware of the SCT that claimed something else.
Relying parties that want to check for inclusion reveal which certificates they are interested in to the log. This makes it possible to infer the browsing history of the requesting party, which violates their privacy.
\cite{blagovStateCertificateTransparency2020,meiklejohnSoKSCTAuditing2022,SCTAuditingGoogle}.

\subsubsection*{SCT Auditing} 

To mitigate the omission attack, Google introduced SCT Auditing \cite{SCTAuditingGoogle}. A dedicated monitor operated by Google receives SCTs from Chrome users who opted in and verifies their inclusion in the logs. It allows this particular monitor to \emph{audit} the log against its issued SCTs. 
\Cref{fig:CT_fullproto} also shows the message flow for SCT Auditing.

\subsubsection*{Current deployment}
\label{sec:current_deployment}

Inspired by known weaknesses of CT, implementations found in browsers deviate from the standard and enforce their own policies. 
We investigated public design documents and tested the implementations in all major browsers in \Cref{tab:browser_CT_impls}. All the policies we encountered define a small set of logs that are trusted and a number of SCTs that are required for a certificate to be accepted.
All designs refer to CTv1 \cite{laurieCertificateTransparency2013}. The newer version, CTv2 \cite{laurieCertificateTransparencyVersion2021}, has not been adopted by any implementation we are aware of, so we will focus on CTv1 in this work.
Notably, no considered browser implements a client-side inclusion proof for privacy and efficiency reasons.

\section{Background}
\label{sec:methodology}

\subsection{Tamarin}

Tamarin performs automated analysis in the Dolev-Yao
model~\cite{meierTAMARINProverSymbolic2013}, wherein
messages are described as abstract terms. 
Terms are composed from uninterpreted function symbols that represent cryptographic primitives and are applied to \emph{variables} representing yet unknown values, as well as \emph{names} serving as the base type, representing either high-entropy values like keys (then the name is \emph{fresh}) or publicly known constants (then the name is \emph{public}).
If we want to clarify the
type, we write $\mathord{\sim}n$ for the former and $\$n$ for the latter.
The behavior of these cryptographic primitives is specified by an
equational theory on terms.
For example, a hash function is represented by a function symbol $\h$ and the empty equational theory, essentially describing it as a random oracle. We write $\h/1$ to indicate that $\h$ takes one parameter.
Message signing and signature verification are typically written as the function symbols
$\sign$, $\verify$, $\ctrue$ and the equation
$\verify(       \sign(m, \sk), m, \pk(\sk)           ) = \ctrue$,
meaning that verifying a correctly signed message reduces to the 0-ary
function symbol (i.e., constant) \ctrue.

\subsubsection{Multiset Rewriting (MSR)}
\label{sec:multiset_rewriting}

The overall state of a protocol execution is described as a multiset
of \emph{facts} where a fact has a fact symbol $f$ and a list of
terms, typically describing the state a protocol party is in. 
For example, $S_2(\$server,\mathord{\sim}nonce)$ if a server with id $\$server$ has previously received a nonce and is ready to respond.

\emph{Rules} then describe the dynamics of the protocol. They have the form
\[ l \msrewrite{a} r \]
where the premises $l$ contain a multiset of facts required to be in
the current state for this rule to be applicable. These are removed
and substituted with those in the conclusion $r$. A set of facts $a$,
the actions, labels this transition.

Persistent facts, prefixed with $!$, will not be consumed. Such facts are, for instance, used to model the adversary's knowledge, which increases over time.
The built-in facts $\In$, $\Out$ and $\Fr$ model network input and output, and the choice of a fresh name, respectively.

\subsubsection{Trace Properties}

Given a set of rules, a protocol execution is any chain of rewrites starting from the empty multiset. 
 A trace is the sequence of labels in an execution, but skipping empty
labels $\emptyset$.
Security properties are expressed in a first-order logic with
quantification over terms and time points (indexes in the trace). 
The atom $f(\ldots)@i$ expresses a fact $f(\ldots)$ that occurs at time
point $i$. The other atoms are term equality and comparison of
time points.

\subsection{Accountability}\label{sec:accountability}

Tamarin has built-in support for formulating and proving accountability~\cite{morioVerifyingAccountabilityUnbounded2021,morioVerifyingAccountabilityUnbounded2021}.
Their definition is protocol-agnostic and based on causality;
a protocol provides accountability for property $\varphi$ if it can
always correctly identify the parties whose deviation from the
protocol (jointly) 
caused~\cite{halpernModificationHalpernPearlDefinition2015} a violation $\neg\varphi$.
Consequently, iff the protocol identifies no such set of parties, then
$\varphi$ must hold, hence accountability for $\varphi$ implies
verifiability for $\varphi$.

In our context, CT's task is to determine whether $\varphi$ was
violated and who was (part of) the cause for that.
Practically, this means that CT specifies that each party provides proof for their own correct behavior and collects evidence of other parties' correct behavior.
Parties might deviate but may be detected by other parties. There is no party with a complete view of the network, which distinguishes accountability in security protocols from
accountability in distributed systems or software dependability.
We consider the punishment of misbehavior to be outside the protocol.
By contrast, the precise conditions under which a party is blamed we
consider very much part of the protocol, and find currently under-specified.
Hence, our task shall be to figure out these conditions (which we call \emph{tests})
and the precise property $\varphi$.

To clarify who provides the inputs and what they indicate,
we impose a naming convention for these events.
$\mathtt{Assert}(\mathit{id},c,s,p)$ means that the party with identity
$\mathit{id}$
makes a claim $c$ regarding $s$ w.r.t.\ a (possibly empty) list 
of parameters $p$. These events shall be emitted by rules modeling
$\mathit{id}$'s role in the protocol and be computable by these
parties.

In Tamarin, tests are defined as trace properties with unbound
variables for one or more party identities.
For instance, the following tests check for a certificate that is
later (externally) found to be incorrect.
\begin{lstlisting}[emph={rootCA}]
test CAfalseAsserts_rootBlamed: 
  "Ex intCA fields ca_fields #t0 #t1 #t2. 
    Asserts('external auditor', 
            'That intermediary CA signed the following incorrect statement and the validation chain says the following about that CA.', 
            intCA, <fields, ca_fields>)@t0
    & Asserts('CCADB', 'No root CA signed this CA.',
              intCA, <ca_fields, rootCA>)@t1"
\end{lstlisting}

%

%
%
Observe that the underlined variable \texttt{rootCA} is unbound. These
so-called \emph{free} variables determine who the test blames.
Given a trace $t$, this test blames $\mathit{ca}$ if the test condition with the free variable 
instantiated to $\mathit{ca}$ holds on $t$, written 
$t \vDash t \set{ \mathtt{rootCA} \mapsto \mathit{ca} }$. 
A test may contain multiple free variables, in which case these
parties are \emph{jointly} blamed. In a mixnet, for instance, an
attack may necessarily involve multiple mixes working together.
But a test can also apply multiple times on the same trace (e.g., if
many attacks are mounted), or multiple tests may be defined and also apply on the same trace. 
In both cases, the instantiations of those tests assign blame sets that are \emph{independently} blamed.
Given a set of tests and a trace $t$, 
the \emph{verdict} is the set of all sets of party identities
for which some test is satisfied on $t$.

For instance, consider a second test below where the external
auditor makes the same assertion, but the root CA correctly documented the
intermediate CA, so the misbehavior must be assigned to the
intermediate CA. %
\begin{lstlisting} [emph={intCA}]
    // [.. as in previous test ...]
    & Asserts(rootCA, 'I signed this intermediate CA.', 
              intCA, <ca_fields>)@t1
\end{lstlisting}
\label{errata_rename_role}

If both tests match, say, for $\mathit{intCA}$ and $\mathit{rootCA}$,
the blame set will be $\set{ \set{\mathit{intCA}}, \set{\mathit{rootCA}}}$,
the responsibility is independent. Indeed, the CCADB, assumed honest,
cannot make the conflicting assertions that the root CA has documented
both the signing CA and has not, so these would be separate instances.

The verdict function induced by the set of traces should be correct,
i.e., it
should identify each minimal set of parties that (jointly) caused an attack (i.e., a violation of $\varphi$) and did so by deviating from their normal behavior (i.e., from the protocol). 
For instance, if $t$ is such that three independent attacks happen, two matching
the first test and one matching the second, then the verdict
identifies three singleton causes, two of them root CAs and one
intermediate CA.
Correctness now asserts that for each attack, the respective
intermediary or root CA was indeed deviating, and without them,
$\varphi$ would not have been violated (in that particular attack).

Morio and Künnemann provide us with a decomposition of accountability
into trace properties, 
each necessary and all
combined sufficient.
This allows automated analysis with Tamarin, but the detailed decomposition is quite involved and not necessary for the understanding of the attacks we find.
We present four of those trace properties that are intuitive and help us intuit attacks that we will see later.
\begin{itemize}
    \item \emph{sufficiency}: Each test has a trace that matches with only the blamed parties corrupted.
    \item \emph{verifiability}: Should no test match, $\varphi$ holds,
        and vice versa.
    \item \emph{minimality:} No strict subset of some test's verdict can trigger that test or any other.
    \item \emph{uniqueness:} Whenever a test matches, the blamed parties must be under adversarial control.
    
\end{itemize}
The remaining two (injectivity and single-matchedness) are never violated in this model and are less intuitive, hence we omit an explanation of them and instead refer the reader to 
\cite{morioVerifyingAccountabilityUnbounded2021}. Tamarin checks them via syntactic
conditions.\footnote{The syntactic conditions are called BR and RP in \cite{morioVerifyingAccountabilityUnbounded2021}.
For BR, all tests in this work succeed. For RP, the syntactic condition in Tamarin is too
strict; we thus checked this condition manually, see
\appendixorfull{sec:replacement-property}.}

\section{The Standard PKI}
\label{sec:pki}

%

We start by laying out the PKI used on the web, which puts unverifiable trust into the CAs.
In the first step we add an external validator and show that it can hold the previous parties accountable but needs itself to be trusted (by each client) and always be online.
Although this is unrealistic, it helps transition to CT, which sets up infrastructure to distribute this validation task through the introduction of dedicated roles.
%
In the following sections, we will proceed similarly for CT and extensions, adding new infrastructure (first monitors, then logs) and showing how trust moves from previously trusted roles to new, typically distributed, infrastructure.
Each time, efficiency improves or parties have better incentives to remain honest.

We use standard function symbols $\pk/1$, $\sign/2$ and $\verify/3$ and the equation
$
    \verify(\sign(x, y), x, \pk(y)           ) = \ctrue()
    $
to model signatures. 
Certificates are signed 7-tuples representing the different fields of the certificate, including the identities of the issuer (1) and subject (2), the subject's public key (3), and the type (4). The remaining fields contain a fresh serial number (5) and a validity period, represented by two timestamps (6 and 7).
In the equation above, $x$ is this 7-tuple, and $y$ is the issuer's secret key.

When the type is set to `selfCert', it represents a trusted root
certificate. Issuer and subject are the same; hence the certificate is
signed with the public key that the tuple contains.
When set to `pubKey-CA' or `pubKey`, it is a regular public key certificate and signed with the issuer's public key, which is different from the subject's (intermediary CA's or end entity) public key.
For `pubKey-CA', the certificate can sign others, i.e., the subject is an intermediate CA. For `pubKey', the certificate is a leaf, i.e., the subject is an end entity, e.g., a web server.

To give a flavor of modeling in Tamarin, we provide the multiset-rewriting (\Cref{sec:multiset_rewriting}) rule that models the client validating an incoming certificate. Here, a CA that signed the certificate is already registered. To validate a certificate, the client checks the following:

\begin{enumerate}
    \item The certificate has been issued and signed by a CA that has been signed by a trusted root CA.
    \item The signature of the intermediate CA certificate verifies under the root CA's public key.
    \item The signature of the certificate verifies under the CA's public key.
    \item The certificate is still valid. 
\end{enumerate}

\begin{lstlisting}[label={lst:client_validate_pki}]
rule ClientValidate:
let
    ca_fields = <$rootCA,$CA,pkCA,'pubKey-CA',
                 snCA,$i0,$j0>
    fields = <$CA,$id,pkID,'pubKey',sn,$i,$j>
    chain = <<fields,pub_sig>,<ca_fields,ca_sig>>
in
[   In(chain), // incoming certificate 
    !Root_CA($rootCA,pkRootCA), // check (1)
] --[ // check (2)+(3)
    Eq(verify(pub_sig,fields,pkCA),true()),
    Eq(verify(ca_sig,ca_fields,pkRootCA),true()),
    // check (4)
    CheckValidUntil($now,$j),
    CheckValidUntil($now,$j0),
    Time($now),
    // emitted event 
    BrowserAcceptsChain($CA,fields,$rootCA,ca_fields),
]-> [ ]
\end{lstlisting}

The action |BrowserAcceptsChain| indicates that this rule was applied, i.e., a client accepted this certificate and the intermediate CA certificate as valid.
We use it to state our security goal $\varphi$.

\subsubsection*{Modeling Timestamps}\label{sec:time_model}

To perform check 4), we need to model the two timestamps included in certificates: time of creation and time of expiration.
Tamarin does not have native support of time beyond ordering actions. 
We use the approach of Morio et al.~\cite{morioAutomatedSecurityAnalysis2023} and model timestamps as public values that are assigned meaning by an axiomatic model. 
Using public variables ensures the adversary can access all timestamps, thereby also simplifying Tamarin's message deduction.

We annotate with an action \lstinline|Time($now)| every rule that needs to measure time.  We relate timestamps with the following restriction. Restrictions are properties that Tamarin assumes to hold.
\begin{lstlisting}
restriction time_monotonicity:
"All t #t1 #t3. Time(t)@t1 & Time(t)@t3 & #t1 < #t3
==> (All tp #t2. Time(tp)@t2 & #t1 < #t2 & #t2 < #t3 
      ==> tp = t )"
\end{lstlisting}
As monotonicity is sufficient for the properties we prove, we avoid other assumptions on timestamps. \texttt{Time} can occur in different parties; hence we assume a global clock.

We check whether an expiry date has passed by putting the action \texttt{CheckValidUntil} in the rule that performs the check. We then constrain traces to those where the check is evaluated correctly:

\begin{lstlisting}
restriction certStillValid:
"All now valid_till #t1.
    Time(now)@t1 & CheckValidUntil(now, valid_till)@t1
    ==> not(Ex #t2. #t2 < #t1 & Time(valid_till)@t2)"
\end{lstlisting}

\subsubsection*{Accountable Authenticity}\label{sec:acc_auth_pki}

\begin{table}
    \begin{tabular}{lp{6cm}}
\toprule
name             & meaning 
\\ 
\midrule 
\textit{fields}           & content of certificate, a 7-tuple \\
\textit{ca\_fields}       & like fields, but type field is `pubKey-CA', typically information about the intermediary CA that signed a leaf certificate \\
\textit{ca\_sig} & signature over ca\_fields, signed with root CA key \\
\textit{ca\_cert }        & the tuple $\langle \mathit{ca\_fields, ca\_sig} \rangle$\\
\textit{CA}      & certificate authority (root or intermediary) \\
\textit{rootCA}  & root CA \\
\textit{L}       & logger                                                                                                                    \\
$\idlog$ & id of a log that is maintained by some logger\\
\textit{M}       & monitor \\
\textit{C}       & client\\
\textit{P} & protocol participant of any role 
\\ \bottomrule
\end{tabular}
\caption{Commonly used variable names.}
\label{tab:variable-names}
\end{table}

\begin{table*}
    \footnotesize
\begin{tabular}{p{50mm}lp{5cm}l}
\toprule
name                                                      & model    & meaning                   & parties providing evidence or assertion \\ \midrule
\textit{BrowserAcceptsChain(CA,fields,rootCA,ca\_fields,$L_1$,$L_2$)}
& all & Browser accepts leaf certificate with
\texttt{fields} signed by \texttt{CA}, who's certificate is signed by
\texttt{rootCA} and with SCTs from $L_1$ and $L_2$. & affected user  
\\
GroundTruth(CA,fields) &
  all &
  Honest domain owner (DO) provided evidence that fields are correct. &
  domain owner \\
ExternalAuditorCheck(ca\_fields, fields) &
  PKI &
  External auditor completed checks for certification chain containing \texttt{ca\_fields} and \texttt{fields}. &
  external auditor, CCADB \\\midrule
  CTCheck(M, fields, CA, ca\_fields)                & CT       
& \multirow{4}{*}{
      \hspace{-2.2em}$\left.\begin{array}{l}
                \\
                \\
                \\
                \\
                \end{array}\right\rbrace $ Monitor completed
            protocol-specific checks.
        }
& monitor, affected client (can be same), CCADB \\
SCTAuditCheck(M, fields, ca\_fields)                     & SCTAudit &                           
& monitor (designated auditor), client, CCADB    \\
SCTReceiptCheck($M$, fields) & SCTAuditReceipt &   
& monitor (designated auditor), client, CCADB  \\ 
STHGossipCheck($M$, $L$, fields, CA, ca\_fields, $s_1$, $s_2$) & STHGossip &   
& monitor, client, CCADB \\ 
CTTransparencyCheck(L, fields)       & CT       
& \multirow{3}{*}{
      \hspace{-2.2em}$\left.\begin{array}{l}
                \\
                \\
                \\
                \end{array}\right\rbrace $ 
                \parbox{40mm}{Monitor completed
                protocol-specific transparency checks.}
        }
& monitor
\\
SCTAuditTr\ldots(L, fields) & SCTAudit &
& monitor (designated auditor), client    \\
STHGossipTr\ldots(M, L, fields, CA, ca\_fields, $s_1$, $s_2$) & STHGossip & 
& monitor (designated auditor), client    \\
\bottomrule
\end{tabular}
\caption{Facts that appear in Lemmas.}
\label{tab:facts}
\end{table*}

We posit that the main goal of the PKI is to provide an authentic mapping from identity to public key, or, put more generally, that when a browser accepts the information in a certificate \texttt{fields} from \texttt{CA}, then \texttt{CA} checked that this information is the ground truth, e.g., by validating the identity and ownership of a domain.

\begin{lstlisting}[label={lst:pki_authenticity}]
lemma authenticity:
"All CA fields #j.
   BrowserAcceptsChain(CA, fields, *, *)@j 
   ==> Ex #i. GroundTruth(CA, fields)@i & #i < #j"
\end{lstlisting}

See \Cref{tab:variable-names} for the naming scheme we use for variables and \Cref{tab:facts} for the facts we will use as predicates in the lemmas.
In particular, \texttt{BrowserAcceptsChain} will obtain more parameters when we introduce CT, so for uniformity, we use `|*|' to signify an arbitrary value, corresponding to all-quantified variables $i_1, i_2, \ldots$ that are otherwise ignored.

We immediately find that a corrupt CA can construct a rogue certificate (\lemmaref{pki}{authenticity}) thereby refuting authenticity.
Tamarin finds two possible corruption scenarios: one where the signing CA itself is compromised, and another where the involved root CA misbehaves by signing a bogus CA certificate which is subsequently used to violate authenticity. A refined version that rules out these two scenarios holds (\lemmaref{pki}{cert_auth}). Both corruption scenarios are possible; we need to be able to hold both individually accountable later.

To motivate CT, assume the user does not trust all root CAs and sometimes wants to validate whether a CA is still honest by externally checking the certificate, e.g., by calling the website owner on the telephone and comparing fingerprints.
We thus add an external validation oracle that obtains a signed certificate and asserts whether its fields are correct or, with a second rule, that they are incorrect.
With an additional restriction, we ensure that the validator never contradicts itself.

Now there is a middle ground between 
unconditional authenticity (refuted in \lemmaref{pki}{authenticity})
and 
outright assuming the (root) CAs are honest (the aforementioned \lemmaref{pki}{cert_auth}). We shall accept that authenticity may be violated, but hold the responsible CA accountable.
Reconsider the tests from \Cref{sec:accountability}, where an external auditor finds incorrect fields within the certificate. As the certificate contains an intermediary CA's signature, the auditor can identify that intermediary CA as responsible (together with the relevant fields in the validation chain).
Root CAs have to publish the intermediate CAs they signed in~\cite{ChromeRootProgram, RootCertificateProgram,MozillaRootStore} into the publicly accessible CCADB database~\cite{CommonCADatabase}.
If the intermediate CA is not in the CCADB, a root CA is to blame, as they signed a rogue intermediate CA without documenting it.
Otherwise, a trusted (by the root CA) intermediary CA misbehaved and caused the attack, so it is to blame.
We can then prove the following lemma:
\begin{lstlisting}[
    caption = {PKI provides acc.\ auth.\ assuming a trusted external
    validator.},
    label = {lst:external_check_acc}]
lemma A_PKI_Ext:
 CAfalseAsserts_rootBlamed, CAfalseAsserts_rootNotBlamed 
 accounts for 
 "All CA rootCA ca_fields fields #t0 #t1.
 ExternalAuditorCheck(ca_fields, fields)@t0
 /*  if the certificate's statement is checked 
 externally... */
 & BrowserAcceptsChain(CA, fields, rootCA, ca_fields)@t1
 ==>
     (Ex #t. GroundTruth(CA, fields)@t)"
     // ...the certificate's statement is correct
\end{lstlisting}

We can use this lemma as a common template for the accountability properties of CT and its extensions. 

\begin{definition}[Accountable authenticity, informal]\label{def:accountable-authenticity} 
We say a model provides accountable authenticity for a check $C$ (represented by a fact symbol) if the above lemma holds with $C$ in place of \texttt{ExternalAuditorCheck}.
\end{definition}
Note that this structure has an implicit liveness assumption: accountability holds only for those fields that were actually checked, i.e., the party performing the check was active. 
Typically, the rule emitting $C(\mathtt{ca\_fields}, \mathtt{fields})$ is triggered by at least two rules, one indicating a violation of accountability and thus triggering some test, and another determining that the evidence for $\mathtt{ca\_fields}$ and $\mathtt{fields}$ does not suggest misbehavior.

We observe that we have shifted trust from the CAs to external validation, which is inefficient. CT distributes this validation task to multiple monitors operating on a distributed log rather than relying on input from the client.
Time to start exploring accountability in CT.

\section{Certificate Transparency}
\label{sec:CT}

Before we add loggers and monitors to the system, we have to amend certificates with a promise by a logger that it will add a certificate soon.

\subsubsection*{Signed Certificate Timestamps (SCTs)}

SCTs consist of three fields: the issuer (usually a logger), a type field, in our case just the type `\texttt{SCT}', and a timestamp to indicate when the SCT was issued.
They represent the promise of the specified logger to add the certificate to the log by the issuing time plus the MMD (typically 24h).
In our model, we modify the certificates to embed two SCTs from distinct logs, as all known browser policies require two SCTs (see \Cref{tab:browser_CT_impls}).
CT-conforming certificates have two embedded SCTs and carry the '\texttt{PubKey}' type. \emph{Unfinished} public key certificates do not embed SCTs and have type '\texttt{pre-cert}', but are otherwise the same.
The signature is calculated over all content fields of the certificate, including embedded SCTs.

\subsubsection*{Monitors and Loggers}

We introduce loggers and monitors with initialization rules.
As the monitor does not have its own key pair, its initialization rule only assigns a public variable, representing its identity. 
Loggers register a public key along with a self-signed certificate. Their key is later used to issue and sign SCTs. The main purpose of logs is to maintain the log structure and hand out commitments (STHs) and proofs for inspecting parties. 
We provide these rules in more detail in \appendixorfull{sec:app-modelling-ct}.

In our model, domain owners can summon a monitor (either themselves or a third party, see \Cref{sec:background}) to check for new certificates whose subject matches the domain owner. We assume the domain owner provides the monitor some \emph{ground truth} on the correctness of the fields in the certificate.
Following the certification guidelines in CT, honest CAs validate the fields during pre-certification.
We thus emit an event $\factsymbol{GroundTruth}$, which we can use to define rogue certificates (and, analogously, benign certificates)~\cite{laurieCertificateTransparency2013}.
\begin{lstlisting}
restriction rogueCert: 
"All CA fields #t1. 
 RogueCert(CA, fields)@t1 ==> 
 not(Ex #t0. GroundTruth(CA,fields)@t0 & #t0 < #t1)"
\end{lstlisting}
Later, the monitor will make use of this knowledge to check for authenticity of certificates it encounters in the logs.

\subsubsection*{Certification}

As discussed in \Cref{sec:background}, certification becomes a two-step process, which we model with two separate rules. The first issues a pre-certificate and sends it out on the network for two loggers to issue SCTs. The second receives these two SCTs and, knowing the pre-certificate, creates a public key certificate that embeds them.
For the client, we adapt the rule \texttt{ClientValidate} from \Cref{lst:client_validate_pki} to additionally verify the two embedded SCTs and add a condition asserting that the two loggers are different. 

\subsubsection*{Modeling Logs}
\label{sec:modeling_logs}

We model each logger's logs via trace properties, similar to how we modeled time. 
The action
\[ \functionsymbol{LoggerComputesAddLog}(L,\idlog, \ldots), \]
e.g., represents additions to the log $\idlog$ maintained by $L$ and is given semantics \emph{axiomatically}, through a restriction on traces.
For the single-ledger case, earlier work~\cite{bruniAutomatedAnalysisAccountability2017a,kunnemannAutomatedVerificationAccountability2019a} followed the same approach but could not represent inclusion proofs in messages on the network, as the log data structure was represented entirely on the trace.
Explicit modeling of the underlying Merkle trees~\cite{chevalAutomaticVerificationTransparency2023} does not scale as well (see \Cref{sec:related-work} for discussion).
Hence, we instead extend the axiomatic approach to handle multiple ledgers, MMD, inclusion proofs, and append-only proofs.
Proofs need to be transmitted on the network and thus require an entirely new axiomatization.
In particular, corrupt loggers must be able to produce `incorrect' logs, so we have a reason to hold them accountable, but if they do, they cannot forge these proofs.
Honest loggers also compute these proofs, but they never break these properties.

Instead of considering the log as the outcome of a series of computations, we record the computations themselves on the log.
Each logger indexes a chain of computations on the logs with an identifier. 
When a logger $L$ adds an entry $f$ to a log $l$, it emits an action $\factsymbol{LoggerComputesAddLog}(L, l, f)$ that indicates a correct addition to the Merkle tree via the function $\mathsf{ADD}$~\cite{crosbyEfficientDataStructures}. 
We can model removal from the log by recomputing it from scratch with a fresh id, but honest loggers maintain only a single log id, as they adhere to CT.
Dishonest parties, however, can compute multiple logs on the fly using all the information they know and thus perform an equivocation.
Any computation that produces a valid Merkle tree is thus available; computations that produce invalid Merkle trees are only represented via malformed proofs, i.e., terms that do not pass validation. Merkle trees are never transmitted, only proofs about them; hence this is w.l.o.g. 

We define predicates $\factsymbol{EntryVisible}$ and $\factsymbol{EntryNotVisible}$, and use them in lemmas and restrictions to define the outcome of a sequence of such computations at a point in time.

\begin{lstlisting}[label={lst:visibility_predicates},mathescape]
EntryVisible(L, f, ca_cert, chain, $\idlog$, #now, sth_ts) <=> 
(Ex sct_ts #t.
/* f was added to L's log $\color{sqBrown}\idlog$ at timepoint in SCT,
   which is the start of MMD period */
   LoggerComputesAddLog(L, $\idlog$, f, ca_cert, 
                        chain, sct_ts)@t 
   & #t < #now
/* ...MMD period has passed */
   & not(sth_ts = sct_ts) 
)
\end{lstlisting}
This rule takes the MMD into account by considering the time of the addition to the log and the time of the snapshot. As a simplification, we consider every action with the same {\lstinline|Time($i)|} fact to belong to the same MMD period and every later timestamp to be after it. In this sense the addition becomes visible when that time has passed.
\factsymbol{EntryNotVisible} is in \appendixorfull{sec:app-modelling-ct}.

\subsubsection*{Inclusion proofs and
snapshots}\label{sec:inclusion-proofs}

We represent inclusion proofs and commitments (STHs) as fresh variables, so they can be linked to the unique time point where they were computed.
We use restrictions to link them to the log at that point in time (the \emph{snapshot}) and give them meaning.
The restrictions assert that iff a proof succeeds (i.e., the corresponding event is triggered on the nonce representing the proof) the proof statement holds true, i.e.,
\begin{itemize}
    \item for an inclusion proof, the entry was visible (see predicate
        |EntryVisible|) at the time the proof was handed out.
    \item for an append-only proof, between two snapshots of a log, no entries disappeared.
\end{itemize}
Merkle trees and proofs constructed with them are thus assumed to work perfectly, which follows Dolev-Yao's `perfect crypto' paradigm but encodes this assumption in the restriction instead of the term algebra.

Proof fetching is a three-step process. When a party, in our model either a client or a monitor, requests a proof,
the logger chooses any snapshot that belongs to them (identified by the public variable $\idlog$) and emits a fresh value  representing the proof for that snapshot, i.e., the log at this moment.
Attackers can present old snapshots of the log by producing a new log (with a new $\idlog$) that contains the entries of the older snapshot.
In the third step, the client or monitor receives this proof and checks for the property of interest. 
We add four restrictions to define the semantics of each proof type and for failure and success. We list and explain them in 
    \processifversion{conf}the full version 
    \fvrefman
        {sec:inclusion_proof,sec:append_only_proof}
        {full:sec:inclusion_proof,full:sec:append_only_proof}%
        .
    Due to space limitations, we can only provide the restriction for a successful inclusion proof: for the branch marked with the event
    \texttt{InclusionProofSucceeds(..)}
the logger indicated in the proof must have presented the view
identified by $\idlog$ prior to the check and the entry whose
inclusion is proven must satisfy the \texttt{EntryVisible} predicate for that time.

\begin{lstlisting}[mathescape=true]
restriction inclusionProofWorks:
"All [...] . 
 InclusionProofSucceeds($\id_\mathit{fetcher}$,L,fields,ca_cert,
                        chain, sth, session)@t5
 ==> Ex id $\idlog$ time #t2. 
  // Logger handed out a snapshot to the recipient..
  LoggerPresentsView(L,id,$\idlog$,time,sth,session)@t2 
  // ...prior to the check
  & #t2 < #t5 
  // and the entry is visible in that view
  & EntryVisible(L,fields,ca_cert,chain,$\idlog$,#t2,time)"
\end{lstlisting}

Note that we assume a private channel for this communication step. This is necessary because otherwise the network attacker could replay an old message from an honest log, producing a failing inclusion test that implicates the (honest!) log.
This attack transfers to the real world, and indeed, CT demands the communication with the logger be implemented via HTTPS GET and POST requests.
Any communication channel with a logger earns authenticity through hard-coded keys of the loggers. Otherwise, we would run into a chicken-and-egg problem, as the authenticity of the logger's public key using CT relies on the loggers themselves. The list of trusted loggers for instance, by Chrome \cite{ChromeKnownCTLogList} includes the public keys of the loggers to be used for pinning. The same applies to the list of trusted loggers by Apple and Mozilla \cite{TrustedLogListApple,MozillaTrustedLogs}. In practice monitoring tools such as CertSpotter rely on these public keys and use them to verify the signed statements by loggers, i.e., STHs \cite{sslmateCertspotterMonitorMonitorgo}.

%
%
%
%
%
%
%


\subsubsection*{Accountable Authenticity}

As discussed before, we assume the monitor knows the ground truth. We thus add rules permitting it to raise an alarm, accept a correct certificate or misbehave by `sleeping', i.e., receiving a certificate but not acting upon it.
Each of these rules produces a fact \textit{MonitorFoundCert} or \textit{MonitorCouldNotFindCert} that is consumed by the actual CT check rules.
These use the browser rules to determine which logs $L_1$ and $L_2$ should hold a certificate (the browser could just be run by the monitor itself) and check the result of the monitor for this certificate on $L_1$ and $L_2$.
Again, this check may succeed or not, so there are two rules that perform it, each emitting the action \textit{CTCheck}.
The \texttt{CTcheck} rule for the case where the monitor returns \textit{MonitorCouldNotFindCert} waits until the MMD has passed and then additionally consults the CCADB so that the following test can determine whether to blame the intermediate CA or the root CA. 
Here, we display the test that blames any documented intermediate CA (\texttt{\underline{CA}}) that issued a certificate which the monitor flagged as rogue.

\begin{lstlisting}[ label={lst:accountable_authenticity_CT}, emph={CA} ]
test ca_auth_on_log:
"Ex M rootCA fields ca_fields #t0 #t1 #t2.
    CTCheck(M, fields, CA, ca_fields)@t0
    & Asserts(M, 'Blame this CA.', CA, <fields, 
      ca_fields>)@t1 // fields contradict ground truth 
    & Asserts(rootCA, 'I signed this intermediate CA.', 
              CA, <ca_fields>)@t2"
\end{lstlisting}

The second test (omitted) does the same, but for undocumented intermediate CAs, thereby blaming any root CA that certified an undocumented intermediate CA.

Although we trust the monitor to confirm the certificate in question, we fail to hold CAs accountable. Tamarin provides a counterexample where a corrupt CA can issue a rogue certificate that is accepted by the client (\ruleref{ct}{ClientAcceptsCert_Key}).
Even if the honest monitor becomes active afterwards and fetches one of the logs that ought to contain the entry for that certificate, a corrupt logger can hide this entry from the monitor.
The monitor would thus fail to blame the CA (let alone the log), and CT would not even ensure verifiability, i.e., that an attack leads to a nonempty blame set and thus at least one test matches.
We can, however, show this lemma if we trust the logger.

\begin{theorem}[
\lemmaref{ct}{A_CT_1},
\lemmaref{ct}{A_CT_2}]
    Against untrusted CAs, but with a trusted monitor and trusted
    logs,
    CTAuditCheck provides accountable authenticity.
    With only trust in the monitor, it does not.
\end{theorem}

\subsubsection*{Transparency}

Given that the log is honest, we have shown that we can prove accountable authenticity.
This motivates our analysis of \emph{transparency} as a separate property.

Ideally, we would demand that, whenever any client accepts a certificate, it can be found in both logs identified by the mandatory two SCTs.
But when precisely is a certificate `in a log'? Loggers are free to equivocate so they can maintain different logs, represented by different $\idlog$, which they may present to different parties---or just keep to themselves, which would render this property useless.

Instead, the idea is that, once an entry has been presented as `in the log', any party checking for this entry will find it.
Again, we define a check, but also a fact marking its positive outcome. When a client accepts a certificate and someone checks for it later, that check must be successful. In the case of \emph{accountable transparency}, if that check is not successful, the parties responsible for failure need to be identifiable. 

\begin{definition}[transparency, informal]\label{def:transparency-informal} 
    A Tamarin model has a transparency check $C$ if the occurrence
    of a fact $C(L,\fields)$ indicates that the check $C$ has been
    conducted for a log $L$ and a certificate statement $\fields$.
    Let furthermore $C^\top$ indicate that the check was successful,
    i.e., the log $L$ confirms at the time of the check to have that
    entry.
    
    For a transparency check $C$, a model $M$ provides \emph{transparency} if
    the following lemma is valid

\begin{lstlisting}[mathescape=true]
All CA fields rootCA ca_fields L1 L2 #t1 #t2.
 BrowserAcceptsChain(CA,fields,rootCA,ca_fields,L1,L2)@t1
  & ( L = L1 | L = L2)
  & #t1 < #t2
  & $C$(L,fields)@t2 ==> $C^\top$(L,fields)@t2 
\end{lstlisting}

    and \emph{accountable transparency}, if a set of tests provides accountability of that property.
\end{definition}

Transparency is also of independent interest to third parties that rely on the correctness of the logs, e.g., researchers~\cite{10.1007/978-3-319-76481-8_13}.
Given CT's name, one could even argue that users might expect this property.

We show that transparency does not hold for dishonest loggers (\lemmaref{ct}{transparency}).
Assuming honest loggers, we find transparency to hold, but only after the MMD (\lemmaref{ct}{transparency_w_MMD_honest_log}).
Observing that honest loggers provide accountability, we can thus focus on the problem of holding the logger accountable.

\section{SCT Auditing}
\label{sec:ctwaudit}\label{sec:auditing}

Chrome adopted \emph{SCT auditing} as an opt-in mechanism in~\cite{OptinSCTAuditing,SCTAuditingGoogle}.
Instead of the clients, the monitors request inclusion proofs from the loggers.
Clients upload a random selection of SCTs and certificate chains they have encountered to a particular version of the monitor currently maintained by Google.
Having a large set of SCTs and underlying certificates at hand puts the monitor in a much stronger position: now, the monitor can validate a logger's entries and past STHs w.r.t.\ those.
If the monitor knows the ground truth, it can use these certificates for authenticity checks. It can also enhance transparency by treating SCT auditing as a secondary source for discovering new certificates, rather than relying solely on logs. Finally, the received SCTs serve as promises that the monitor can verify.
Any SCTs that do not have a counterpart in the log the monitor retrieves indicate misbehavior after the MMD has elapsed. 

The downside to this proposal is that clients that opted in expose parts of their browsing history, as the certificates they emit contain the websites they visited as the subject.

\subsubsection*{Modeling}

We add a rule for clients to report the certificates they encountered to a trusted monitor. 
As the mechanism is opt-in, the rule needs not be used.
Accordingly, the properties below talk about the guarantees for participating clients, but consider a model where not every client does that.
We add rules to the monitor that use incoming certificates and perform an authenticity check on them, emitting \texttt{SCTAuditCheck}.

As in the previous versions, the monitor relies on the ground truth it has received from the actual domain owner for this. There are two possible outcomes of this check, modeled in a rule each: 
Either the monitor finds a certificate inconsistent with its ground truth, or the certificate is authentic. Moreover, a dishonest monitor can find a rogue certificate but \emph{sleep on it}, meaning they neither inform the subject nor blame the CA. It is necessary to distinguish this case from the monitor not receiving the certificate, as SCT audit relies on reception of the message, yet a misbehaving monitor can merely make a false claim (which we ignore in the model, as the domain owner can promptly verify it to be false) or ignore the received certificate.

\subsubsection*{Accountable Authenticity}\label{sec:acc_auth_CTAudit}

We list the complete set of tests for accountable authenticity here. 
As before, the tests distinguish, based on the CCADB, whether we blame the root CA or the intermediate CA, hence there are two.
Both include the event $\texttt{SCTAuditCheck}$, thus precluding that the client (browser) opted into this check.

\begin{lstlisting}[emph={CA}]
test ca_rogue_audit:
 "Ex M ca_fields fields rootCA #t #t0 #t1.
  SCTAuditCheck(M, fields, ca_fields)@t
  & Asserts(M, 'This CA has the following fields and it failed the SCT audit for the following fields.', CA, 
            <fields, ca_fields>)@t0
  & Asserts(rootCA, 'I signed this intermediate CA.', CA, 
            <ca_fields>)@t1
 "
\end{lstlisting}

\begin{lstlisting}[emph={rootCA}]
test root_ca_rogue_audit:
 "Ex M ca_fields fields CA #t #t0 #t1. 
  SCTAuditCheck(M, fields, ca_fields)@t
  & Asserts(M, 'This CA has the following fields and it failed the SCT audit for the following fields.', CA, 
            <fields, ca_fields>)@t0
  & Asserts('CCADB', 'No root CA signed this CA.', CA, 
            <ca_fields, rootCA>)@t1
 "
\end{lstlisting}

The monitor identifies misbehaving CAs correctly, even without trust in the loggers. Note, though, that the monitor is assumed to have ground truth.

\begin{theorem}[\lemmaref{sct_audit}{A_CTAudit_1}]\label{lst:monitor_external_check_acc_audit}
    Against untrusted loggers and CAs, but with a trusted monitor,
    SCTAuditCheck provides accountable authenticity. 
\end{theorem}

If we do \emph{not} trust the monitor, we find a counterexample where the monitor receives the rogue certificate but \emph{sleeps on it} (ignores it), so the audit does not happen (\lemmaref{sct_audit}{A_CTAudit_2}).

SCT Auditing mitigates the threat from malicious loggers hiding log entries.
Having that \emph{all clients} audit is a strong assumption, but a probabilistic check was actually deployed in Chrome.
The above lemma is easily raised to the probabilistic setting, as it specifies its guarantees per audited certificate, e.g., if Chrome clients audit a certificate with probability $p$ and $n$ rogue certificates are sent to such clients, then an attack carries the risk of detection with a probability of $1 - (1-p)^n$.

We also trust that the monitor performs the audit and knows the ground truth. 
This was a realistic assumption in previous scenarios, where companies could run their own monitors. 
Here, however, monitors receive certificates from all clients that have opted in, which is (a) a substantial load and (b) prohibitively slow for the clients if the monitor is not fast
enough or has high latency. 
Finally, (c), if each server were trusted to also act as a monitor, then we would have eliminated the need for a PKI, let alone CT, altogether.
It is thus not surprising that Chrome's monitor is run by Google and trusted by users running Google's web browser Chrome.
Anyone can become a monitor, but not every monitor can perform SCT auditing. 

\subsubsection*{Transparency}

Uniquely for SCT auditing, the monitor can use the inclusion check to hold loggers accountable, too.
This is useful for third parties that rely on the log.
Chrome's design documentation hints that, indeed, such a check might exist. 
\begin{quote}
`Google will [..] raise an alert if any log misbehavior is detected. The server-side auditing and alerting is covered in a separate Google-internal doc.' \cite{OptinSCTAuditing}
\end{quote}

We find that transparency still fails (\lemmaref{sct_audit}{transparency_SCT_Auditing}) but that we can provide accountable transparency, holding dishonest loggers accountable for failure to provide transparency.

\begin{lstlisting}[emph={L}]
test audit_inclusionViol:
"Ex M fields session #t #t1.
  Asserts(M, 'This loggers log currently does not include these fields.', L, <fields, session>)@t
& SCTAuditTransparencyCheckSession(L, fields, session)@t1
"
\end{lstlisting}

When the monitor performs an audit (which now requires it to obtain the STH externally, e.g., from a client), it fetches a log view and runs the usual inclusion check. If the monitor cannot find the certificate, the logger is blamed. With this procedure, we only need to trust the monitor, given that the client that provided the STH verified the inclusion of the certificate in that STH. 

\begin{theorem}[\somethingref{sct_audit}{T_Audit}]
    Against untrusted loggers and CAs, but with a trusted monitor,
    \textit{SCTAuditTransparencyCheck} provides accountable
    transparency. 
\end{theorem}
 
\subsubsection*{Holding the Monitor Accountable With Receipts}

So far, we have excluded monitor misbehavior from our accountability analysis, as CT does not provide us with a mechanism suited for accountability tests. Other parties cannot verify whether the monitor purposefully ignores a rogue certificate or just did not receive it.

To see if that is possible at all, we extend SCT auditing, the only model managing to achieve accountable authenticity with a dishonest logger.  We add a rule that has monitors hand over a signed receipt to the client, confirming the monitor has seen a particular snapshot of the log. Clients can require such a receipt before they proceed.
In case of a dispute or suspicion, another rule has the client forward this receipt to the domain owner (alternatively, the domain owner could act as the client for testing).

In case a rogue certificate is in active use and found by the domain owner, an additional test can detect the monitor's misbehavior.
While at that point, the domain owner could now just as well serve as an external validator like in~\cref{sec:acc_auth_pki}, the point here is that they can oversee the monitor service, which they contracted for this purpose.
Moreover, SCT auditing works best if one monitor receives many SCTs and certificates, which would entail that this is an external service and independent enough that trust becomes an issue.
Combined with the previous tests, we can now show accountable authenticity in the case where 
CAs, loggers, and monitors can be corrupted (\lemmaref{ct_receipt}{A_Receipt}).

The model thus demonstrates a mechanism to hold monitors accountable as well. But note that it does not represent a published proposal, nor is it thought out at that level of detail.

\section{STH Gossiping}
\label{sec:ctwgossip}\label{sec:gossiping}

We now turn to a different, yet similar approach to the equivocation problem, STH gossiping.
While in SCT auditing, certificates had to be shared with the monitor to validate SCTs for the inclusion property, 
STHs can be validated for the append-only property without the certificate.
This means STHs can be shared more widely: 
Gossiping clients can share STHs with each other and verify append-only-ness between the snapshots they received and their own.
To attack transparency, the logger must hide the entry from every gossiping client it interacts with, which prevents partition attacks on the gossiping clients.
Direct communication is, however, impractical, since there are no standard communications method between browsers and end users are often difficult to address via IP because they are behind home routers or firewalls.

Who should the STH then be gossiped to? \textcite{nordbergGossipingCT2018} presents various options, which can be classified by where the gossip is finally received and then analyzed. These are either the auditor or the monitor. 
We model specifically the case where STHs are sent to the monitor
(comparable to \emph{STH Pollination}~\cite{nordbergGossipingCT2018})  or, equivalently, where the monitor doubles as a trusted auditor
(\emph{Trusted Auditor Relationship}~\cite{nordbergGossipingCT2018}).
Compared to \emph{STH Pollination} (or \emph{SCT Feedback}~\cite{nordbergGossipingCT2018}, which is the same but additionally for SCTs), we assume direct communication between client and monitor, instead of a relay via the web server, which was implemented as an HTTP header~\cite{10.1007/978-3-319-76481-8_13,ExpectCTHeaderHTTP2025} but is now deprecated. As we do not trust the web server, this abstraction comes without loss of generality. Our model thus abstracts both STH~Pollination and Trusted Auditor Relationship, depending on whether we call the receiving monitor a `trusted auditor' or not.

\subsubsection*{Modeling}

We add a new rule that broadcasts the client's STH to the monitor.
The monitor performs authenticity checks similar to those in CT, using the entries on the log to find potentially rogue certificates. In fact, we can reuse the same rules for authenticity checks in STH gossiping.
What is new, however, is that the monitor receives a gossiped STH from a client. For this STH, inclusion of a specific entry may have already been proven by the client.
To make use of this information, we add new rules to check append-only-ness between the gossiped STH and a freshly fetched snapshot from the log. So far, we have only considered append-only checks between snapshots that have been locally stored by the monitor or client, respectively. In this variant, one such STH is received from the network instead.

In \appendixorfull{sec:app-modelling-ct-gossiping} we provide the modified rules for this. They emit an assertion if the append-only check between the gossiped STH and the newly fetched snapshot fails.
We assume the monitor to be honest and reuse its authenticity checks on the log from \Cref{sec:CT}. If the certificate contradicts the known key for its subject, it blames the CA.

\subsubsection*{Accountable authenticity}\label{sec:acc_auth_CTGossip}

The monitor fetches a new snapshot from the log in order to retrieve the same certificate and perform an authenticity check with it.
If the certificate contradicts the ground truth, it will assert that the CA is rogue; otherwise, it proceeds silently. As before, we use this assertion and the previous legitimacy checks for intermediate CAs to decide to either blame the root CA or the intermediate CA.

Still, accountable authenticity does not hold. We find an attack against verifiability (\lemmaref{sct_audit}{A_Gossip_1}). A malicious logger alone can bypass the test by first sending the client a correct certificate and an STH that proves inclusion, but then creating a new, incorrect, refutable inclusion proof for a different snapshot that does not contain this certificate.
That proof will fail and the monitor will realize that something is wrong. It even correctly blames the logger, as discussed in the next section.
But without the actual certificate, the monitor cannot decide if the CA is involved or not, or even whether authenticity was actually violated by whatever certificate the client accepted.
Put differently, by creating an incorrect inclusion proof, the monitor can give rise to the impression that a certificate is missing and thus produce a false alarm, even if that certificate is completely benign. 
So, in contrast to SCT Auditing, sharing STHs with monitors is not sufficient to achieve accountable authenticity against untrusted loggers, but is sufficient for the monitor to detect equivocation.
Gossiping ultimately achieves similar results to plain CT w.r.t. accountable authenticity, but with the added benefit of detecting equivocation by the logger (accountable transparency).

\begin{theorem}[
Lemmas \modelhrefplus{ct_gossip}{A_Gossip_0}
and
\somethingref{ct_gossip}{A_Gossip_1}]
    Against untrusted CAs, but with trusted loggers and a trusted monitor,
    \emph{STHGossipCheck} provides accountable
    authenticity.
\end{theorem}

\subsubsection*{Transparency}

The successful identification of the equivocating logger in the test for accountable authenticity gives us hope for (accountable) transparency.

Starting optimistically, we find that straightforward transparency is violated by a misbehaving logger
(\lemmaref{ct_gossip}{transparency_Gossip}).
Assume that a client has a successful inclusion proof for a certificate in a given STH and gossips this STH with the monitor.
The logger may now refuse to disclose the entries behind that fetched STH to the monitor and instead equivocate.
This breaks transparency to the monitor, even though the client enjoys transparency.

However, a monitor can detect this situation if it additionally checks for the append-only property. We derive a test that achieves accountable transparency rather than transparency
(\lemmaref{ct_gossip}{T_Gossip}).
Summarizing rules 
\modelhref{ct_gossip}{GossipTransparencyCheck_Success}{\path{Gossip}\-\path{TransparencyCheck_Success}}
and
\modelhref{ct_gossip}{GossipTransparencyCheck_Failed}{-\path{_Failed}},
the trans\-pa\-rency check is run by the monitor for each target logger $L$. The monitor tests, for each gossiped STH it receives whether
    the most recent STH it fetched from the log is
        append-only w.r.t. to the gossiped STH
and
\emph{assumes} that
    the client from which the gossip originates from has checked for
        inclusion---this is a modeling assumption stating that
        sufficiently many clients participate.

\begin{theorem}[
    \modelhrefplus{ct_gossip}{transparency_Gossip}
    and
\somethingref{ct_gossip}{T_Gossip}]
    Against untrusted loggers and CAs, but with a trusted monitor,
    \emph{STHGossipTransparencyCheck} provides accountable
    transparency, but does not provide transparency. 
\end{theorem}

\section{Results}
\label{sec:evaluation}

\begin{table*}
\centering
\begin{tabular}{@{}llllll@{}}
\toprule
protocol & \multicolumn{4}{c}{authenticity with trust in \ldots} & transparency \\
\cmidrule(l){2-5} 
\cmidrule(l){6-6} 
         & no one & monitor & + logger & + ext.\ validator & (no trust) \\
\midrule
PKI            & --- & --- & --- 
               &  \omark~\plref{pki}{A\_PKI\_Ext}
               \plnotref{pki}{authenticity}
               & ---
\\

CT             & \xmark~\plref{ct}{A\_CT\_1} 
               & \xmark~\plref{ct}{A\_CT\_2} 
               & \omark~\plref{ct}{A\_CT\_3}\plnotref{ct}{authenticity}
               & \omark~$\implies$
               & \xmark~\plref{ct}{T_CT_1}
\\

SCT Auditing   & 
               \xmark \plref{sct_audit}{A\_CTAudit\_2}
               & \omark~\plref{sct_audit}{A\_CTAudit\_1}\plnotref{sct_audit}{authenticity} 
               & \omark~$\implies$ 
               & \omark~$\implies$
               & \omark~\plref{sct_audit}{T\_Audit} 
\\
\quad + receipts   & 
\omark~\plref{ct\_receipt}{A\_Receipt}\plnotref{ct\_receipt}{authenticity}
               & \omark~$\implies$
               & \omark~$\implies$
               & \omark~$\implies$
               & \omark~$\implies$
\\

Gossiping      & \xmark 
               & \xmark~\plref{ct_gossip}{A\_Gossip\_1}
               & \omark~\plref{ct_gossip}{A\_Gossip\_2}\plnotref{ct_gossip}{authenticity}
               & \omark~$\implies$
               & \omark~\plref{ct_gossip}{T\_Gossip}
\\
\bottomrule
\multicolumn{6}c{
(\xmark = attack \quad \omark = acc. auth. / transp. \quad  \cmark = property proven \quad --- = not applicable)
}
\\
\end{tabular}
\caption{
Results: Accountability and Transparency in CT. 
Click the \plref{pki}{} icon to jump to the lemma in the browser.
Accountable authenticity is implied by authenticity, so we
additionally provide a counterexample against authenticity with the
icon \plnotref{pki}{}. 
A $\implies$ denotes that a lemma
follows logically (e.g., when adding more trust or for
straight-forward extensions like adding receipts).
}
\label{tab:accountability_summary}
\end{table*}

    

We summarize our results and provide details about our approach and the proof effort.
All results were computed on an Intel 13th Gen Core i7-13700H with 16 allocated cores and 14 GB of memory.
The proof files are available at \supplementURL; links to lemmas in this document are clickable and lead to anchors in an HTML document.

\subsubsection*{Sanity Lemmas}

To show our models internally consistent, we show that each rule is reachable, thus ensuring that security statements are non-vacuous. 
We prove 192 such lemmas, one for each rule. Each is proven automatically, but we sometimes add restrictions to limit the search space (only for sanity lemmas and other lemmas that check for satisfiability, e.g., sufficiency).
Their total verification time is less than two hours.


\subsubsection*{Protocol mechanics}

To convince ourselves that we understand the mechanics of the protocol, we proved authenticity properties that found the trust between parties. 
Tamarin takes only a few minutes to prove all of them.

\begin{lemma}[ \lemmaref{pki}{cert_auth}]
    When a client accepts a certificate, then the CA asserts the value
    of this certificate unless this particular CA was corrupted.
\end{lemma}
\begin{lemma}[ \lemmaref{ct}{cert_sct_auth}] 
        If a CA asserts a public key certificate with two SCTs, 
        then two distinct loggers have previously issued them honestly
        unless some party was corrupted.
\end{lemma}
\begin{lemma}[ \lemmaref{ct}{sct_auth}] 
            If a client accepts a certificate with two SCTs,
            then the two loggers they refer to have added an entry to
            their log unless one of them is corrupt. 
\end{lemma}


\subsubsection*{Accountability}

We summarize the results discussed in the previous sections in
\Cref{tab:accountability_summary}. 
The overall verification time for all automated lemmas (sanity, accountability,
transparency and helping lemmas) across all models is around five hours.
For three helping lemmas and three accountability lemmas, we required
partial manual intervention.


\subsubsection*{Transparency}

\Cref{tab:accountability_summary} confirms the intuition that transparency is essential for achieving accountable authenticity---unless we can trust the logger.
To hold CAs accountable, log entries must include the full certificate, including the fields signed by the CA. Without them, loggers may be able to accuse CAs wrongly. 
More generally, \emph{what data} the protocol makes transparent determines who
can be held accountable: full certificates enable accountability of CAs, SCTs suffice to hold loggers accountable for omissions.
Observe that SCT auditing covers both.
STH gossiping, however, reveals STHs to monitors, but not CA signatures, so CAs cannot be held accountable.
This explains the gap: while STH gossiping can provide accountable transparency, which could be a building block for accountable authenticity, it fails to achieve accountable authenticity because it can only blame loggers correctly, but not CAs.

\subsubsection*{Applicability to Current Web Browsers}\label{sec:design_eval}

In \Cref{sec:browser_impls}, we discussed how different browsers implement CT and extensions.
SCT auditing, implemented in Chrome, leads to strong transparency and (accountable) authenticity guarantees, but only if we assume every
certificate is audited.
In practice, not every Chrome client is configured to do this; those that are only audit only a randomized portion of certificates~\cite{OptinSCTAuditing}.
Nevertheless, we can state that every handshake that is actually audited enjoys these guarantees.

The other browsers neither implement SCT auditing nor provide client-side inclusion proofs. 
As we have seen, client-side inclusion proofs by themselves do not guarantee transparency---and thus accountability without trust in at least one logger---as a dishonest logger can equivocate. 
%
But they could be exploited to more effect. To avoid getting caught equivocating, the logger has to prove inclusion to clients but hide the entry from the monitor. Depending on the implementation, a logger can distinguish a client who requests an inclusion proof from a monitor who instead fetches all entries. If monitors mimicked the client's behavior by first fetching the STH and then requesting inclusion proofs for all known certificates, this distinction would be harder. We know existing implementations are not designed to do that, as clients are typically not communicating with the logger at all, except Chrome's designated SCT auditing monitor maintained by Google itself. 

As of now, authenticity and transparency still rely on trust in the loggers.
All browsers require $n=2$ SCTs from distinct loggers. These can be different for each certificate, but must come from a pool of $N=7$ or $N=8$ trusted loggers (see \Cref{tab:browser_CT_impls}).
Assuming the attacker knows the user's browser, they can pick $n$ dishonest loggers specifically. 
Hence, the trust assumption is that at least $N-(n-1)$ loggers are trusted (or at least, not colluding).
Currently $N-(n-1)$ equals 6 or 7, depending on the browser. The logger pool is almost the same, so an attack would be largely independent of the browser.

\section{Related Work}
\label{sec:related-work}

So far, formal analyses of CT have either been conducted manually, in the cryptographic standard model~\cite{wrotniakProvableSecurityPKI2024}, or with automated tools in the Dolev-Yao model, but with significant simplifications~\cite{chevalAutomaticVerificationTransparency2023,morioVerifyingAccountabilityUnbounded2021}.

\subsubsection*{Cryptographic analyses of CT}\label{sec:herzberg_acc}

\textcite{wrotniakProvableSecurityPKI2024} modeled CT and conducted manual analysis to show transparency and accountability.
They define accountability as a 23-line-long game in which the attacker wins if they are able to produce a rogue certificate that validates, although neither the root CA was corrupted nor any CA on the chain (including the root CA) was responsible for signing the rogue certificate's fields.
They show that this notion of accountability holds for CT and PKIX.

First, this cryptographic game is not only specific to accountability for PKI schemes but also to their formulation, and encodes protocol-specific concepts like root certificates and root CAs, specific fields in certificates, and the data flow of the scheme. This definition is hard to adapt to formal analysis outside the cryptographic model, as it is game-based and very complex. 

Secondly, because the game requirement is so specific, it is not clear whether it achieves accountability in a broader sense. 
Compared with the definition we employ, their definition at most guarantees that some responsible party can be identified.
It does not ensure that this party indeed bears responsibility nor that everyone else who should bear responsibility is identified. 
Indeed, there is no mechanism to blame parties others than CAs or to blame more than one.
Many works on accountability in protocol-like settings highlight the ability to blame (all) the responsible participants~\cite{feigenbaumOpenVsClosed2014,chocklerResponsibilityBlameStructuralmodel2003,kustersAccountabilityDefinitionRelationship2010a}. 

By contrast, our definition of accountability was tested on OCSP Stapling and Mixnets~\cite{morioVerifyingAccountabilityUnbounded2021} and can ensure all parties blamed are causes of the violation and all such causes are indeed blamed.
\cite{kustersAccountabilityDefinitionRelationship2010a} employs protocol-specific policies to this end.
For this reason, we can find that accountability only holds under strong assumptions.
Nevertheless, \cite{wrotniakProvableSecurityPKI2024}
motivated the present analysis and helped formulate some tests (see below).

\subsubsection*{Automated Analysis of Accountability in CT}

CT was first formally verified for accountability by \textcite{bruniAutomatedAnalysisAccountability2017a}. \textcite{morioVerifyingAccountabilityUnbounded2021} extend the model and analyze it with respect to the same accountability definition we use here, pointing out that their definition may miss violations and might blame a party for a violation of a security property that cannot be violated.
We do not further extend their model but instead obtain a broader and much more realistic model that captures the details of the actual specification while continuing to support an unbounded number of participants.

\textcite{chevalAutomaticVerificationTransparency2023} analyzed CT (but no extensions) in ProVerif. 
Their focus is on accurately modeling the Merkle tree structure. They extend ProVerif to support an axiomatic modeling of ledgers. Besides CT, ledgers are also used in voting systems
\cite{cortierFiftyShadesBallot2020}.
They first show the Merkle tree structure secure and then, assuming a Merkle tree interface, that a monitor receiving a proof can be sure that the certificate is in a (single, centralized) log, but not other properties. 
\begin{full}
For instance, they model the proof of inclusion (they call this a proof of presence) using inductive clauses. They allow computing the proof recursively, starting from the leaf that contains the entry we are interested in, whose proof would be the empty list. Then, depending on whether the right or left child node is missing, the hash of that node is appended to the proof list together with a label \texttt{left} or \texttt{right}, indicating the branch taken to traverse the tree from the root to the leaf we started at.
\end{full}
Our trace-based model of the log (\Cref{sec:CT}) also allows for equivocation and represents proofs as messages that can be sent around but builds on restrictions that axiomatically assert the properties of the log in relation to inclusion or append-only proofs.
We do not prove those axioms.
%
Despite the more detailed modeling of Merkle trees, their model of CT is simpler than ours in arguably more pertinent aspects: they only consider browser-side validation (in a simplified manner) and browser-side monitoring (which is not implemented by any popular browser). Neither the loggers nor the monitors appear as parties distinct from the client or website and hence cannot be individually corrupted.
In summary, for accountability, we had to design our model from scratch. Moreover, we are the first to model time to represent the certificate validity period and represent the pre-certification process.

\subsubsection*{SCT Auditing} \label{sec:sct_auditing}

\textcite{wrotniakProvableSecurityPKI2024}'s model also considers SCT auditing~\cite[CTwAudit, Sec. D.2]{wrotniakProvableSecurityPKI2024} (see \Cref{sec:auditing}). There, SCT auditing is given as an algorithm that, given a certificate with SCTs, returns the identity of the loggers that are deemed responsible for the lack of transparency should the corresponding entry not be visible in the log.
They show that auditing can be used to achieve "audited transparency", i.e., either transparency holds or they are able to blame the responsible logger if transparency is violated.
Similarly, we can show accountable transparency (for STH gossiping) or even transparency itself (for SCT auditing) under lighter assumptions.
Our accountability tests for transparency in STH gossiping are similar to their algorithm for audited transparency, but they strictly focus on transparency with respect to the log.
Instead of using the audited certificate to learn an entry, their algorithm `magically' hands a certificate to the monitor, who has a snapshot from \emph{every} log, and compares it with the existing knowledge. We do not assume that a monitor stores every certificate it has ever seen, but instead, it fetches another snapshot from a log and blames them if inclusion is violated.

\section{Conclusion}

We discussed the authenticity problem in PKIs that rely on a central authority. Our results support that CT solves this problem and successfully delegates the required trust from CAs to loggers. This helps; we gave the Symantec CA misbehavior as an example in the introduction.
%
By logging the whole certificate and chain, log entries provide enough evidence to hold CAs accountable for the certificates they issued.
Corrupt loggers can hide evidence, which constitutes a known attack against accountable authenticity and transparency.
If one of the loggers in use is honest, transparency holds, but only for that particular logger.
An honest monitor that knows the ground truth---every domain owner can be a monitor---can hold the CA accountable for maliciously issued certificates. 

Assuming an honest logger is not as benign as it looks and requires further research from the policy perspective. The current pool of loggers consists of 7--8 loggers (see \Cref{tab:browser_CT_impls}) with almost\footnote{Currently, IPnG is included in some but not all.} complete intersect.
Given Chrome's high market share, and the fact that certificates are transmitted before the user agent is known, there is no incentive for websites to include additional loggers, let alone exclude loggers from Chrome's list.
Essentially, Google decides which set of loggers is trusted, even for other browsers. Google's loggers are part of this set. 

With SCT auditing, implemented in Chrome, we can remove the log from the trust assumption and instead hold misbehaving loggers and CAs simultaneously accountable.
We find, however, that this strong guarantee only holds for certificates that are audited; for other certificates, loggers can get away with hiding the corresponding entry.
Certificates are only audited randomly and only if the client uses Chrome and has opted in, thereby sharing them with one specific, trusted Google monitor. 
We discussed the privacy implications in \Cref{sec:auditing}.
At the policy level, it is worthwhile to explore how to counteract the inherent tendency for centralization in SCT auditing.

With STH gossiping, which is deprecated and was not implemented, clients also share their log snapshot, but they additionally perform an inclusion proof with the corresponding logger.
This adds further cost and effectively eliminates the privacy advantage gained from sharing only the log snapshot instead of full certificates. 
Future work could explore privacy-preserving primitives to compute inclusion proofs.
Accountable transparency holds; while loggers can refuse to show the entries corresponding to an STH they have handed out, the monitor can identify this misbehavior.
The same attack violates accountable authenticity.

Our model offers significant improvements compared to existing work, most notably the approach to Merkle proofs and its coverage of dishonest intermediary CAs, root CAs, loggers, and monitors.
It currently does not cover certificate revocation.
While CT does not oblige logs to store revocation requests, researchers are considering integrating revocation proofs into the CT log, e.g., via postcertificates~\cite{korzhitskiiPostcertificatesRevocationTransparency2022} or similar entries~\cite{kongCTngSecureCertificate2021} stored in the log that can prove a certificate was presented after being revoked.
Vice versa, if revocation works, it also helps CT: if a monitor detects a rogue certificate, the most immediate reaction should be to revoke it.
Accountability is as necessary for revocation as it is for certificate authenticity: 
\textcite{morioVerifyingAccountabilityUnbounded2021} showed that OCSP Stapling (RFC 6066) can provide accountability, but only if we trust the OCSP responder.
Future work should investigate possible synergies between certificate transparency and revocation.


\printbibliography

@inproceedings{10.1007/978-3-319-76481-8_13,
  title = {In Log We Trust: {{Revealing}} Poor Security Practices with Certificate Transparency Logs and Internet Measurements},
  booktitle = {Passive and Active Measurement},
  author = {Gasser, Oliver and Hof, Benjamin and Helm, Max and Korczynski, Maciej and Holz, Ralph and Carle, Georg},
  editor = {Beverly, Robert and Smaragdakis, Georgios and Feldmann, Anja},
  date = {2018},
  pages = {173--185},
  publisher = {Springer International Publishing},
  location = {Cham},
  isbn = {978-3-319-76481-8}
}

@inproceedings{amannMissionAccomplishedHTTPS2017,
  title = {Mission Accomplished?: {{HTTPS}} Security after Diginotar},
  shorttitle = {Mission Accomplished?},
  booktitle = {Proceedings of the 2017 {{Internet Measurement Conference}}},
  author = {Amann, Johanna and Gasser, Oliver and Scheitle, Quirin and Brent, Lexi and Carle, Georg and Holz, Ralph},
  date = {2017-11},
  pages = {325--340},
  publisher = {ACM},
  location = {London United Kingdom},
  doi = {10.1145/3131365.3131401},
  url = {https://dl.acm.org/doi/10.1145/3131365.3131401},
  urldate = {2026-03-16},
  eventtitle = {{{IMC}} '17: {{Internet Measurement Conference}}},
  isbn = {978-1-4503-5118-8},
  langid = {english}
}

@online{ApplesCertificateTransparency2025,
  title = {Apple's {{Certificate Transparency}} Policy},
  date = {2025-04-21},
  url = {https://support.apple.com/en-us/103214},
  urldate = {2025-05-27},
  langid = {english},
  organization = {Apple Support}
}

@article{BaselineRequirementsIssuance2025,
  title = {Baseline {{Requirements}} for the {{Issuance}} and {{Management}} of {{Publicly-Trusted TLS Server Certificates}}},
  date = {2025-05-16},
  journaltitle = {CA/Browser Forum},
  publisher = {CA/Browser Forum},
  langid = {english}
}

@article{blagovStateCertificateTransparency2020,
  title = {State of the {{Certificate Transparency Ecosystem}}},
  author = {Blagov, Nikita},
  namea = {Architectures, Chair Of Network},
  nameatype = {collaborator},
  date = {2020},
  publisher = {{Chair of Network Architectures and Services, Department of Computer Science, Technical University of Munich}},
  doi = {10.2313/NET-2020-11-1_09},
  url = {https://www.net.in.tum.de/fileadmin/TUM/NET/NET-2020-11-1/NET-2020-11-1_09.pdf},
  urldate = {2025-06-30},
  langid = {english}
}

@incollection{bruniAutomatedAnalysisAccountability2017a,
  title = {Automated {{Analysis}} of {{Accountability}}},
  booktitle = {Information {{Security}}},
  author = {Bruni, Alessandro and Giustolisi, Rosario and Schuermann, Carsten},
  editor = {Nguyen, Phong Q. and Zhou, Jianying},
  date = {2017},
  volume = {10599},
  pages = {417--434},
  publisher = {Springer International Publishing},
  location = {Cham},
  doi = {10.1007/978-3-319-69659-1_23},
  url = {https://link.springer.com/10.1007/978-3-319-69659-1_23},
  urldate = {2025-04-02},
  isbn = {978-3-319-69658-4 978-3-319-69659-1},
  langid = {english}
}

@online{CAEntrustIssues,
  title = {{{CA}}/{{Entrust Issues}} - {{MozillaWiki}}},
  url = {https://wiki.mozilla.org/CA/Entrust_Issues},
  urldate = {2026-03-16}
}

@online{chevalAutomaticVerificationTransparency2023,
  title = {Automatic Verification of Transparency Protocols (Extended Version)},
  author = {Cheval, Vincent and Moreira, José and Ryan, Mark},
  date = {2023-04-16},
  eprint = {2303.04500},
  eprinttype = {arXiv},
  eprintclass = {cs},
  doi = {10.48550/arXiv.2303.04500},
  url = {http://arxiv.org/abs/2303.04500},
  urldate = {2025-03-17},
  pubstate = {prepublished}
}

@online{chocklerResponsibilityBlameStructuralmodel2003,
  title = {Responsibility and Blame: A Structural-Model Approach},
  shorttitle = {Responsibility and Blame},
  author = {Chockler, Hana and Halpern, Joseph Y.},
  date = {2003-12-17},
  eprint = {cs/0312038},
  eprinttype = {arXiv},
  doi = {10.48550/arXiv.cs/0312038},
  url = {http://arxiv.org/abs/cs/0312038},
  urldate = {2025-04-08},
  langid = {english},
  pubstate = {prepublished}
}

@online{ChromeCertificateTransparency,
  title = {Chrome {{Certificate Transparency Policy}}},
  url = {https://googlechrome.github.io/CertificateTransparency/ct_policy.html},
  urldate = {2025-05-27},
  langid = {american},
  organization = {CertificateTransparency}
}

@online{ChromeKnownCTLogList,
  title = {Known {{CT}} Logs},
  author = {Chrome},
  url = {https://www.gstatic.com/ct/log_list/v3/log_list.json},
  urldate = {2026-08-02}
}

@online{ChromeRootProgram,
  title = {Chrome {{Root Program Policy}}, {{Version}} 1.7},
  url = {https://googlechrome.github.io/chromerootprogram/#2-common-ca-database},
  urldate = {2025-11-02}
}

@online{CommonCADatabase,
  title = {Common {{CA Database}} by the {{Linux Foundation}}},
  url = {https://www.ccadb.org/policy#5-audit-disclosures},
  urldate = {2025-10-14}
}

@inproceedings{cortierFiftyShadesBallot2020,
  title = {Fifty {{Shades}} of {{Ballot Privacy}}: {{Privacy}} against a {{Malicious Board}}},
  shorttitle = {Fifty {{Shades}} of {{Ballot Privacy}}},
  booktitle = {2020 {{IEEE}} 33rd {{Computer Security Foundations Symposium}} ({{CSF}})},
  author = {Cortier, Véronique and Lallemand, Joseph and Warinschi, Bogdan},
  date = {2020-06},
  pages = {17--32},
  issn = {2374-8303},
  doi = {10.1109/CSF49147.2020.00010},
  url = {https://ieeexplore.ieee.org/document/9155128/},
  urldate = {2025-07-10},
  eventtitle = {2020 {{IEEE}} 33rd {{Computer Security Foundations Symposium}} ({{CSF}})}
}

@online{ctprojectMonitorsCertificateTransparencyn.d.,
  title = {Monitors : {{Certificate Transparency}}},
  year = {\bibstring{nodate}},
  url = {https://certificate.transparency.dev/monitors/},
  urldate = {2025-07-09}
}

@online{ExpectCTHeaderHTTP2025,
  title = {Expect-{{CT}} Header - {{HTTP}} | {{MDN}}},
  date = {2025-06-23},
  url = {https://developer.mozilla.org/en-US/docs/Web/HTTP/Reference/Headers/Expect-CT},
  urldate = {2025-06-30},
  langid = {american}
}

@inproceedings{feigenbaumOpenVsClosed2014,
  title = {Open vs. Closed Systems for Accountability},
  booktitle = {Proceedings of the 2014 {{Symposium}} and {{Bootcamp}} on the {{Science}} of {{Security}}},
  author = {Feigenbaum, Joan and Jaggard, Aaron D. and Wright, Rebecca N.},
  date = {2014-04-08},
  pages = {1--11},
  publisher = {ACM},
  location = {Raleigh North Carolina USA},
  doi = {10.1145/2600176.2600179},
  url = {https://dl.acm.org/doi/10.1145/2600176.2600179},
  urldate = {2025-04-04},
  eventtitle = {{{HotSoS}} '14: {{Symposium}} and {{Bootcamp}} on the {{Science}} of {{Security}}},
  isbn = {978-1-4503-2907-1},
  langid = {english}
}

@online{FirefoxReleaseNotes,
  title = {Firefox},
  url = {https://www.firefox.com/en-US/firefox/android/145.0/releasenotes/},
  urldate = {2026-04-18},
  langid = {english},
  organization = {Firefox for Android 145.0 Release Notes}
}

@online{halpernModificationHalpernPearlDefinition2015,
  title = {A {{Modification}} of the {{Halpern-Pearl Definition}} of {{Causality}}},
  author = {Halpern, Joseph Y.},
  date = {2015-05-01},
  eprint = {1505.00162},
  eprinttype = {arXiv},
  eprintclass = {cs},
  doi = {10.48550/arXiv.1505.00162},
  url = {http://arxiv.org/abs/1505.00162},
  urldate = {2025-07-10},
  langid = {english},
  pubstate = {prepublished}
}

@online{kongCTngSecureCertificate2021,
  title = {{{CTng}}: {{Secure Certificate}} and {{Revocation Transparency}}},
  shorttitle = {{{CTng}}},
  author = {Kong, Jie and James, Damon and Leibowitz, Hemi and Syta, Ewa and Herzberg, Amir},
  date = {2021},
  number = {2021/818},
  url = {https://eprint.iacr.org/2021/818},
  urldate = {2025-10-13},
  pubstate = {prepublished}
}

@online{korzhitskiiPostcertificatesRevocationTransparency2022,
  title = {Postcertificates for {{Revocation Transparency}}},
  author = {Korzhitskii, Nikita and Nemec, Matus and Carlsson, Niklas},
  date = {2022-03-03},
  eprint = {2203.02280},
  eprinttype = {arXiv},
  eprintclass = {cs},
  doi = {10.48550/arXiv.2203.02280},
  url = {http://arxiv.org/abs/2203.02280},
  urldate = {2025-10-13},
  pubstate = {prepublished}
}

@online{kunnemannAutomatedVerificationAccountability2019a,
  title = {Automated {{Verification}} of {{Accountability}} in {{Security Protocols}}},
  author = {Künnemann, Robert and Esiyok, Ilkan and Backes, Michael},
  date = {2019-05-08},
  eprint = {1805.10891},
  eprinttype = {arXiv},
  eprintclass = {cs},
  doi = {10.48550/arXiv.1805.10891},
  url = {http://arxiv.org/abs/1805.10891},
  urldate = {2025-03-18},
  langid = {english},
  pubstate = {prepublished}
}

@inproceedings{kustersAccountabilityDefinitionRelationship2010a,
  title = {Accountability: Definition and Relationship to Verifiability},
  shorttitle = {Accountability},
  booktitle = {Proceedings of the 17th {{ACM}} Conference on {{Computer}} and Communications Security},
  author = {Küsters, Ralf and Truderung, Tomasz and Vogt, Andreas},
  date = {2010-10-04},
  pages = {526--535},
  publisher = {ACM},
  location = {Chicago Illinois USA},
  doi = {10.1145/1866307.1866366},
  url = {https://dl.acm.org/doi/10.1145/1866307.1866366},
  urldate = {2025-04-08},
  eventtitle = {{{CCS}} '10: 17th {{ACM Conference}} on {{Computer}} and {{Communications Security}} 2010},
  isbn = {978-1-4503-0245-6},
  langid = {english}
}

@report{laurieCertificateTransparency2013,
  title = {Certificate {{Transparency}}},
  author = {Laurie, B. and Langley, A. and Kasper, E.},
  date = {2013-06},
  number = {6962},
  pages = {RFC 6962},
  institution = {RFC Editor},
  doi = {10.17487/RFC6962},
  url = {https://www.rfc-editor.org/info/rfc6962},
  langid = {english}
}

@article{laurieCertificateTransparency2014,
  title = {Certificate Transparency},
  author = {Laurie, Ben},
  date = {2014-09-23},
  journaltitle = {Communications of the ACM},
  shortjournal = {Commun. ACM},
  volume = {57},
  number = {10},
  pages = {40--46},
  issn = {0001-0782, 1557-7317},
  doi = {10.1145/2659897},
  url = {https://dl.acm.org/doi/10.1145/2659897},
  urldate = {2026-03-16},
  langid = {english}
}

@report{laurieCertificateTransparencyVersion2021,
  title = {Certificate {{Transparency Version}} 2.0},
  author = {Laurie, B. and Messeri, E. and Stradling, R.},
  date = {2021-12},
  number = {9162},
  pages = {RFC9162},
  institution = {RFC Editor},
  doi = {10.17487/RFC9162},
  url = {https://www.rfc-editor.org/info/rfc9162},
  langid = {english}
}

@incollection{meierTAMARINProverSymbolic2013,
  title = {The {{TAMARIN Prover}} for the {{Symbolic Analysis}} of {{Security Protocols}}},
  booktitle = {Computer {{Aided Verification}}},
  author = {Meier, Simon and Schmidt, Benedikt and Cremers, Cas and Basin, David},
  editor = {Sharygina, Natasha and Veith, Helmut},
  editora = {Hutchison, David and Kanade, Takeo and Kittler, Josef and Kleinberg, Jon M. and Mattern, Friedemann and Mitchell, John C. and Naor, Moni and Nierstrasz, Oscar and Pandu Rangan, C. and Steffen, Bernhard and Sudan, Madhu and Terzopoulos, Demetri and Tygar, Doug and Vardi, Moshe Y. and Weikum, Gerhard},
  editoratype = {redactor},
  date = {2013},
  volume = {8044},
  pages = {696--701},
  publisher = {Springer Berlin Heidelberg},
  location = {Berlin, Heidelberg},
  doi = {10.1007/978-3-642-39799-8_48},
  url = {http://link.springer.com/10.1007/978-3-642-39799-8_48},
  urldate = {2025-07-02},
  isbn = {978-3-642-39798-1 978-3-642-39799-8},
  langid = {english}
}

@article{meiklejohnSoKSCTAuditing2022,
  title = {{{SoK}}: {{SCT Auditing}} in {{Certificate Transparency}}},
  shorttitle = {{{SoK}}},
  author = {Meiklejohn, Sarah and DeBlasio, Joe and O’Brien, Devon and Thompson, Chris and Yeo, Kevin and Stark, Emily},
  date = {2022-07},
  journaltitle = {Proceedings on Privacy Enhancing Technologies},
  shortjournal = {PoPETs},
  volume = {2022},
  number = {3},
  pages = {336--353},
  issn = {2299-0984},
  doi = {10.56553/popets-2022-0075},
  url = {https://petsymposium.org/popets/2022/popets-2022-0075.php},
  urldate = {2025-03-18},
  langid = {english}
}

@online{MerkleTown2025,
  title = {Merkle {{Town}}},
  date = {2025},
  url = {https://ct.cloudflare.com/},
  urldate = {2025-07-20}
}

@inproceedings{morioAutomatedSecurityAnalysis2023,
  title = {Automated {{Security Analysis}} of {{Exposure Notification Systems}}},
  shorttitle = {978-1-939133-37-3},
  booktitle = {32nd {{USENIX Security Symposium}} ({{USENIX Security}} 23)},
  author = {Morio, Kevin and Esiyok, Ilkan and Jackson, Dennis and Künnemann, Robert},
  date = {2023-08},
  pages = {6593--6610},
  publisher = {USENIX Association},
  url = {https://www.usenix.org/conference/usenixsecurity23/presentation/morio},
  langid = {english}
}

@inproceedings{morioVerifyingAccountabilityUnbounded2021,
  title = {Verifying {{Accountability}} for {{Unbounded Sets}} of {{Participants}}},
  booktitle = {2021 {{IEEE}} 34th {{Computer Security Foundations Symposium}} ({{CSF}})},
  author = {Morio, Kevin and Künnemann, Robert},
  date = {2021-06},
  pages = {1--16},
  publisher = {IEEE},
  location = {Dubrovnik, Croatia},
  doi = {10.1109/CSF51468.2021.00032},
  url = {https://ieeexplore.ieee.org/document/9505190/},
  urldate = {2025-03-18},
  eventtitle = {2021 {{IEEE}} 34th {{Computer Security Foundations Symposium}} ({{CSF}})},
  isbn = {978-1-7281-7607-9},
  langid = {english}
}

@online{mozillaRevokingTrustTwo2013,
  title = {Revoking {{Trust}} in {{Two TurkTrust Certificates}}},
  author = {{Mozilla}},
  date = {2013-01-03},
  url = {https://blog.mozilla.org/security/2013/01/03/revoking-trust-in-two-turktrust-certficates},
  urldate = {2025-10-09},
  langid = {american},
  organization = {Mozilla Security Blog}
}

@online{MozillaRootStore,
  title = {Mozilla {{Root Store Policy}}},
  url = {https://www.mozilla.org/en-US/about/governance/policies/security-group/certs/policy/},
  urldate = {2025-11-02},
  langid = {english},
  organization = {Mozilla}
}

@online{MozillaTrustedLogs,
  title = {Known {{CT Logs}}},
  author = {Mozilla},
  url = {https://hg-edge.mozilla.org/mozilla-central/file/tipsecurity/ct/CTKnownLogs.h},
  urldate = {2026-08-02},
  langid = {american}
}

@report{nordbergGossipingCT2018,
  type = {Internet Draft (Expired)},
  title = {Gossiping in {{CT}}},
  author = {Nordberg, Linus and Gillmor, Daniel Kahn and Ritter, Tom},
  date = {2018-01-14},
  number = {draft-ietf-trans-gossip-05},
  institution = {Internet Engineering Task Force},
  url = {https://datatracker.ietf.org/doc/draft-ietf-trans-gossip},
  urldate = {2025-04-02},
  pagetotal = {57}
}

@online{obrienChromesPlanDistrust2017,
  title = {Chrome’s {{Plan}} to {{Distrust Symantec Certificates}}},
  author = {O’Brien, Devon and Sleevi, Ryan and Whalley, Andrew},
  date = {2017-11-09},
  url = {https://security.googleblog.com/2017/09/chromes-plan-to-distrust-symantec.html},
  urldate = {2025-07-15},
  langid = {english},
  organization = {Google Online Security Blog}
}

@online{OptinSCTAuditing,
  title = {Opt-in SCT Auditing (public)},
  author = {Stark, Emily and Thompson, Chris},
  url = {https://docs.google.com/document/d/1G1Jy8LJgSqJ-B673GnTYIG4b7XRw2ZLtvvSlrqFcl4A},
  urldate = {2026-03-17},
  langid = {ngerman},
  organization = {Google Docs}
}

@online{RootCertificateProgram,
  title = {Root {{Certificate Program}} - {{Apple}}},
  url = {https://www.apple.com/certificateauthority/ca_program.html},
  urldate = {2025-11-02}
}

@report{SCTAuditingGoogle,
  title = {{{SCT Auditing Google Source}}},
  institution = {Google},
  url = {https://chromium.googlesource.com/chromium/src/+/refs/heads/main/services/network/sct_auditing/}
}

@online{SecurityEngineeringCertificateTransparency,
  title = {{{SecurityEngineering}}/{{Certificate Transparency}} - {{MozillaWiki}}},
  url = {https://wiki.mozilla.org/SecurityEngineering/Certificate_Transparency},
  urldate = {2025-04-02}
}

@online{sleeviSustainingDigitalCertificate2015,
  title = {Sustaining {{Digital Certificate Security}}},
  author = {Sleevi, Ryan},
  date = {2015-10-28},
  url = {https://security.googleblog.com/2015/10/sustaining-digital-certificate-security.html},
  urldate = {2025-07-15},
  langid = {english},
  organization = {Google Online Security Blog}
}

@online{sslmateCertspotterMonitorMonitorgo,
  title = {Certspotter/Monitor/Monitor.Go at Master · {{SSLMate}}/Certspotter},
  author = {{SSLMate}},
  url = {https://github.com/SSLMate/certspotter/blob/master/monitor/monitor.go},
  urldate = {2026-08-14},
  langid = {english},
  organization = {GitHub}
}

@report{TLS-RFC8446,
  title = {The {{Transport Layer Security}} ({{TLS}}) {{Protocol Version}} 1.3},
  author = {Rescorla, E.},
  date = {2018-08},
  institution = {Internet Engineering Task Force (IETF)},
  issn = {2070-1721},
  url = {https://datatracker.ietf.org/doc/html/rfc8446}
}

@online{TLSPolicy2024,
  title = {{{TLS Policy}}},
  date = {2024-12-11},
  url = {https://github.com/brave/brave-browser/wiki/TLS-Policy},
  urldate = {2025-05-27},
  langid = {english},
  organization = {GitHub}
}

@online{TrustedLogListApple,
  title = {Current {{CT Log List}}},
  author = {Apple},
  url = {https://valid.apple.com/ct/log_list/current_log_list.json},
  urldate = {2026-08-02}
}

@online{UserAgentsCertificaten.d.,
  title = {User {{Agents}} : {{Certificate Transparency}}},
  year = {\bibstring{nodate}},
  url = {https://certificate.transparency.dev/useragents/},
  urldate = {2025-07-03}
}

@inproceedings{wrotniakProvableSecurityPKI2024,
  title = {Provable {{Security}} for {{PKI Schemes}}},
  booktitle = {Proceedings of the 2024 on {{ACM SIGSAC Conference}} on {{Computer}} and {{Communications Security}}},
  author = {Wrótniak, Sara and Leibowitz, Hemi and Syta, Ewa and Herzberg, Amir},
  date = {2024-12-02},
  pages = {1552--1566},
  publisher = {ACM},
  location = {Salt Lake City UT USA},
  doi = {10.1145/3658644.3670374},
  url = {https://dl.acm.org/doi/10.1145/3658644.3670374},
  urldate = {2025-03-19},
  eventtitle = {{{CCS}} '24: {{ACM SIGSAC Conference}} on {{Computer}} and {{Communications Security}}},
  isbn = {979-8-4007-0636-3},
  langid = {english}
}

@misc{acc-ct-supplement,
    author = "Treitz, Timo and Künnemann, Robert",
    title = {Accountability in Certificate Transparency and Variants},
    url = "https://github.com/rkunnema/formal-model-certificate-transparency/blob/main/README.md",
}

@inproceedings{crosbyEfficientDataStructures,
author = {Crosby, Scott A. and Wallach, Dan S.},
title = {Efficient data structures for tamper-evident logging},
year = {2009},
publisher = {USENIX Association},
address = {USA},
booktitle = {Proceedings of the 18th Conference on USENIX Security Symposium},
pages = {317–334},
numpages = {18},
location = {Montreal, Canada},
series = {SSYM'09}
}

\appendix
\crefalias{section}{appendix}



\section{Ethics considerations}

We perform a defense-focussed analysis on the specification
level.
No experiments were conducted on living subjects.

The potential impact of this work is to provide clarity over the
existing guarantees that CT provides, and what structural changes can
improve those. Moreover, new suggestions that may improve the
transparency and accountability properties, but may incur a
little network overhead.
Attacks we find pertain to protocols whose accountability guarantees
were so far not clearly communicated, so they should be considered
boundary conditions of our positive results.

\section{Open Science}

We provide in the supplementary material~\cite{acc-ct-supplement}:

\begin{itemize}
    \item the model files
    \item manually derived proofs
    \item a reproducible environment based on Nix flakes to rerun the
        experiments, including a fixed version of tamarin and
        batch-tamarin for convenient display of results
    \item recorded proof and evaluation results for the reviewer's
        convenience
    \item marked up html version of the model files that the macros in
        this file refer to
    \item a \texttt{README.md} file that explains how the material is
        organized and how the experiments are reproduced.
\end{itemize}

We plan to submit these model files to the Tamarin main repository
eventually as well
as to a long-term archival website.




    \section{Replacement Property}\label{sec:replacement-property}

For every accountability lemma the Replacement Property (RP) must be manually verified when the model uses public constants or restrictions.
Because the model uses both extensively, we provide a brief argument that RP is not violated in our model.
The property states that for every single-matched trace of a case test, the instantiation of free parties can be replaced by any other party which is permitted by our model.
Our model uses public constants as type identifiers for different certificates.
By design, these public constants are not used as identifiers for different roles and for each role initialization, we restrict that the used public variable is unique among different roles.
We need to argue that all used restrictions preserve the replacement property. All inlined restrictions are concerned with asserting function symbol outcomes, distinctness of SCTs and the loggers involved in them. Two more inline restrictions state that the two loggers considered in SCTs must be distinct. They remain distinct under bijective renaming.
All restrictions that axiomatically assert the Merkle proofs and content delivered by a logger are independent of role identifiers, and thus they do not restrict the identities of involved parties.
One additional restriction determines whether the CA documentation is valid for any party role and whether a certificate is valid or benign in an external check.
We restrict CAs, loggers and domain owners to register only once to ensure uniqueness of identities such that the public variable cannot be reused for different roles, including monitors. These eight restrictions remain distinct under bijective renaming.
Three more restrictions concern the uniqueness of log identifiers. Recall that these log identifiers are distinct from the identifiers of loggers. 
The remaining restrictions concern equality, time axioms, and consistent assertions used in the external check. None of them influence role identifiers.
We use two more restrictions in the additional receipt model to decide whether a monitor is blamed. This holds for any monitor.

Thus our restrictions and use of public constants do not violate RP and our analysis results are valid.

\begin{full}
    \section{Modeling CT (Details)}\label{sec:app-modelling-ct}

\subsection{Adding Loggers and Monitors}

We introduce loggers and monitors similarly to the other roles. Both roles have an initialization rule. 
As the monitor does not have its own key pair, it only consists of a public variable representing its identity. 
Loggers register a public key that is later used to issue and sign SCTs. The main purpose of logs is to maintain the log structure and hand out commitments and proofs for inspecting parties. 
The log model is described in detail are given in \fvref{sec:log_structure}.

\begin{lstlisting}
rule Monitor_Init:
 [  ] 
 --[ Time($now), Monitor_Registered($M) ]-> 
 [ !Monitor($M) ]

rule Logger_Init:
 [ Fr(~skL) ] 
 --[ Time($now), Logger_Registered($L) ]->
 [ !Logger_s($L, ~skL, pk(~skL)), !Logger($L, pk(~skL)) ]
\end{lstlisting}

We model the monitoring check by adding a rule that represents a monitor starting to observe a logger and a second rule that represents the start of tracking a specific subject. 

\begin{lstlisting}
rule StartMonitoring:
 [ !Monitor($M), !Logger($L, pkL) ]
 --[ Time($now) ]->
 [ !LoggerDict($M, $L) ]

rule monitorTrackSubject:
 [ !Monitor($M), !DomainOwner(su, pubK) ]
 --[ Time($now) ]->
 [ !TrackSubject($M, su) ]
\end{lstlisting}

\subsection{Modified Certification}
\begin{lstlisting}
rule PreCertify: /* CA issues a pre-certificate */
let
 preC_fields = 
    <$CA, $id, pkID, 'pre-cert', ~sn, $now, $j>
 preC_sig = sign(preC_fields, skCA)
 preC = <preC_fields, preC_sig>
in
 [ 
  !DomainOwner($id, pkID),
  !CA_s0($CA, skCA, pkCA, ca_fields, ca_sig), Fr(~sn) ]
 --[ 
  GroundTruth($CA, 
              <$CA, $id, pkID, 'pubKey', ~sn, $now, $j>),        
  Time($now)
 ]->
 [ Certificate_Store($CA, preC), Out(preC_fields), 
   Out(preC_sig), Out(ca_fields), Out(ca_sig) ]

rule PubKeyCertify:
let 
 preC_fields = 
    <$CA, $id, pkID, 'pre-cert', sn, $i, $j>
 sct1_fields = 
    <$L1, 'SCT', $sct1, <$CA, $id, pkID, sn, $i, $j>>
 sct2_fields = 
    <$L2, 'SCT', $sct2, <$CA, $id, pkID, sn, $i, $j>>
 sct1 = <$L1, 'SCT', $sct1, sct1_sig>
 sct2 = <$L2, 'SCT', $sct2, sct2_sig>
 pub_fields = 
    <$CA, $id, pkID, 'pubKey', sn, $i, $j, sct1, sct2>
 fields = <$CA, $id, pkID, 'pubKey', sn, $i, $j>
 pub_sig = sign(pub_fields, skCA)
in 
[ !CA_s0($CA, skCA, pkCA, ca_fields, CA_sig), 
  /* 2 SCTs: */  In(sct1), In(sct2), 
  !Logger($L1, pkL1), !Logger($L2, pkL2),
  Certificate_Store($CA, <preC_fields, sig_preC>)
] 
--[ 
  Eq(verify(sct1_sig, sct1_fields, pkL1), true()), 
  Eq(verify(sct2_sig, sct2_fields, pkL2), true()), 
  Eq(verify(sig_preC, preC_fields, pkCA), true()), 
  _restrict( not($L1 = $L2) ),
  CheckValidUntil($now, $j),
  Time($now)
]-> 
[ Out(pub_fields), Out(pub_sig) ]
\end{lstlisting}

\subsection{Log Data Structure}\label{sec:log_structure}

\paragraph{Using the trace as ledger}
We model logs maintained by every logger exclusively as a property of the trace.
We add actions to rules to make additions to the log, as well as the generation and verification of proofs visible. Then we use trace restrictions to enforce relations between these. This is similar to the model of time and space in \cite{morioAutomatedSecurityAnalysis2023}.
For example, the action \lstinline|LoggerComputesAddLog($LogIdentity, ...)| represents additions to the log maintained by \lstinline|$LogIdentity|.

A similar approach has been used before by \textcite{bruniAutomatedAnalysisAccountability2017a} and extended by \textcite{kunnemannAutomatedVerificationAccountability2019a} to model CT's public ledgers. Both capture the partition attack by allowing the logger to present different snapshots to different auditors. Inclusion or append-only proofs were not part of either models, and we show that the approach can be further extended to abstractly capture properties of the Merkle tree structure, including append-only proofs and inclusion proofs.

An honest logger accepts an incoming pre-certificate only if it has a full chain that ends with a known root certificate. There are different notions in literature whether the root certificate itself is considered a part of the chain. In our model, the root CA certificate is not part of the chain as every participant in the system has access to it anyway.
Then, it issues the corresponding SCT and adds the chain and pre-certificate to its publicly visible log. Following \cite{laurieCertificateTransparency2013}, the entry in the Merkle tree later (which we refer to interchangeably as log) only contains the pre-certificate without the signature. Note that this has practical reasons, as later with inclusion proofs, the requesting party needs to specify which entry it wants to receive a proof of inclusion. The party might not have the full chain available, nor is it guaranteed that the exact same chain is stored in the log (think of corruption cases). The promise by the log that clients receive as SCT is only computed over the pre-certificate fields and not a chain. Thus we can only hold a logger liable for the pre-certificate fields and not the complete chain.
However, \cite{laurieCertificateTransparency2013} also specifies that loggers MUST store the full chain for auditing purposes. Our action fact \lstinline{LoggerComputesAddLog}, thus contains both the pre-certificate fields and the full chain. Inclusion proofs are later formalized over the pre-certificate fields while monitors inspecting the full log will also get access to the full chain.

\begin{lstlisting}[mathescape=true]
rule add_preC_to_log:
let
 preC_fields = <$\texttt{\$}$CA, $\texttt{\$}$id, pkID, 'pre-cert', sn, $\texttt{\$}$i, $\texttt{\$}$j> 
 ca_fields = 
    <$\texttt{\$}$rootCA, $\texttt{\$}$CA, pkCA, 'pubKey-CA', snCA, $\texttt{\$}$i0, $\texttt{\$}$j0>
 fields = <$\texttt{\$}$CA, $\texttt{\$}$id, pkID, 'pubKey', sn, $\texttt{\$}$i, $\texttt{\$}$j>
 sct_fields = 
    <$\texttt{\$}$L, 'SCT', $\texttt{\$}$now, <$\texttt{\$}$CA, $\texttt{\$}$id, pkID, sn, $\texttt{\$}$i, $\texttt{\$}$j>>
 sct_sig = sign(sct_fields, skL)
 sct = <$\texttt{\$}$L, 'SCT', $\texttt{\$}$now, sct_sig>
 chain = <<preC_fields, preC_sig>, <ca_fields, ca_sig>>
in
[ 
  !Root_CA($\texttt{\$}$rootCA, pkRootCA), !Log($\texttt{\$}$L, $\idlog$),  
  /* Hardcoded key material of the root CA */
  !Logger_s($\texttt{\$}$L, skL, pk(skL)), 
  /* Logger key material (known to relying parties) */
  In(preC_fields), In(preC_sig), 
  In(ca_fields), In(ca_sig)
  /* incoming pre-cert chain */
]
--[
  Eq(verify(preC_sig, preC_fields, pkCA), true()), 
    /* honest additions have to be properly signed */
  Eq(verify(ca_sig, ca_fields, pkRootCA), true()),
  LoggerComputesAddLogFields($\texttt{\$}$L, $\idlog$, fields, $\texttt{\$}$now),
  LoggerComputesAddLog($\texttt{\$}$L, $\idlog$, fields, 
                       <ca_fields, ca_sig>, chain, $\texttt{\$}$now),
  LoggerAsserts($\texttt{\$}$L, sct),
  
  CheckValidUntil($\texttt{\$}$now, $\texttt{\$}$j),
  Time($\texttt{\$}$now),
]->
[ Out(sct) ]
\end{lstlisting}

\subsection{Corrupted Loggers}
\label{sec:corrupted_logging}

Like for CAs, we add a key leakage rule for the logger. This allows the network attacker to construct forged SCTs. To allow manipulation of the log entries, we add rules that can only be used by corrupted loggers.

A maintained log provides three properties that can be checked by every participant:

\begin{enumerate}
    \item Append-only: No entries disappear from the log over time.
    \item SCT Inclusion: Issued SCTs are backed by an entry in the log.
    \item Entry Validity: Every entry in the log corresponds to a valid pre-certificate with a full chain ending in a trusted root.
\end{enumerate}

We add the ability to break precisely these properties in any possible way. Every logger can maintain arbitrarily many logs; each log can be described by the operations of addition and proof construction.

This snapshot is obtained by adding to every operation on the log (addition, proof construction) a $\idlog$, representing the log on which the action is computed.
Whenever some party inspects the entries, the logger controls which log it represents.
Honest loggers maintain just one log, i.e., there is an injection from \lstinline|$IdL| to $\idlog$.

\begin{lstlisting}[mathescape=true]
rule Start_Log_malicious:
 [ !Corrupted_Logger($\texttt{\$}$L, ~skL, pk(~skL)) ]
 --[ StartedLog($\texttt{\$}$L, $\idlog$), Time($\texttt{\$}$now) ]->
 [  ]

rule forge_log_view: 
let
 preC_fields = <$\texttt{\$}$CA, $\texttt{\$}$id, pkID, 'pre-cert', sn, $\texttt{\$}$i, $\texttt{\$}$j> 
 fields = <$\texttt{\$}$CA, $\texttt{\$}$id, pkID, 'pubKey', sn, $\texttt{\$}$i, $\texttt{\$}$j>
 ca_fields = 
    <$\texttt{\$}$rootCA, $\texttt{\$}$CA, pkCA, 'pubKey-CA', snCA, $\texttt{\$}$i0, $\texttt{\$}$j0>
 chain = <<preC_fields, preC_sig>, <ca_fields, ca_sig>>
in
[ !Corrupted_Logger($\texttt{\$}$L, skL, pk(skL)), In(preC_fields), 
  In(preC_sig), In(ca_fields), In(ca_sig),
  /* Incoming certificate that is added */ ]
--[
  /* observe that we can have arbitrary forged additions: 
  there is no requirement for proper signing */
  LoggerComputesAddLogFields($\texttt{\$}$L, $\idlog$, fields, $\texttt{\$}$now),
  LoggerComputesAddLog($\texttt{\$}$L, $\idlog$, fields, 
                       <ca_fields, ca_sig>, chain, $\texttt{\$}$now),
  Time($\texttt{\$}$now)
]->
[ ]
\end{lstlisting}

The forge log entry rule allows corrupted loggers to add arbitrary entry to any of their logs without possessing a valid pre-certificate chain.

\subsection{Visibility Predicates and Maximum Merge Delay}

\paragraph{Maximum Merge Delay (MMD)}
In CT, loggers are allowed to have a time period between SCT issuance and the corresponding entry becoming visible in the log, called maximum merge delay (MMD).
We include this notion in our model by modifying additions to the log and changing the visibility predicates.
Additions to the log now become scheduled at the time they are performed. The fact is annotated with the current model time according to the |Time| action fact. Scheduled additions are pending and, in the sense of MMD, become visible to inspecting parties after the delay has passed. 

\paragraph{Visibility of Merkle Tree Leaves}

To formulate that an entry is visible at a given time point in the trace, we introduce visibility predicates that can be reused for this purpose.
Intuitively we say that an entry is visible in the considered log with $\idlog$ if at the given time point on the trace, the entry has been added (|LoggerComputesAddLog|) before. Likewise, an entry is not visible if it has never been added. We formalize visibility as predicates based on the |#time| of snapshot creation.

\begin{lstlisting}[label={lst:visibility_predicates-full}, mathescape=true]
EntryVisible(L, fields, ca_cert, chain, $\idlog$, #time, 
             sth_time) <=> 
 (Ex different_time #t.
  LoggerComputesAddLog(L, $\idlog$, fields, ca_cert, 
                       chain, different_time)@t 
    /* there has been an addition to $\color{sqBrown}\idlog$ with the 
       fields and chain... */
  & #t < #time
    /* ...prior to #time and... */
  & not(sth_time = different_time) 
    /* ...that addition was visible */
 )

EntryMissing(L, fields, $\idlog$, #time, sth_time) <=>
 not(Ex ca_cert different_time chain #t. 
    LoggerComputesAddLog(L, $\idlog$, fields, ca_cert, 
                         chain, different_time)@t
    & not(sth_time = different_time) 
     /* or there has never been a visible addition 
        to the log */
    & #t < #time
    )
\end{lstlisting}

We include the time of the proof handout (|proof_time|) in the predicate. Additions to the log are visible at |proof_time|, if they have been scheduled to a strictly |different_time|, which represents a different MMD interval. 
Removals from the log are assumed to become immediately visible when they occur in the trace.

|EntryNotVisible| is adapted similarly: either the entry has not been \emph{visibly} added before, or it has been visibly added but removed after and not visibly added after the removal at the time of the inspection.

In the following, we will assume these modified visibility predicates and the setting with MMD.

Note here that these predicates are not concerned with the chain. The visibility predicate is exclusively a property defined on the Merkle tree of the log. We will reuse these restrictions to formulate inclusion proofs on that Merkle tree. Following \cite{laurieCertificateTransparency2013}, an inclusion proof alone does not provide the recipient with any guarantees on the chain validity or existence.

\paragraph{Visibility of Chains}

A monitor that inspects a log is usually interested in more than just a proof of inclusion of a pre-certificate. Think of the case where a monitor finds a potentially rogue pre-certificate entry included in the Merkle tree structure. Without a chain of certificates it is impossible to prove which CA issued the certificate and thus who is to blame.
For this case, we adapt the visibility predicates to capture the pre-certificate and its chain together.

\begin{lstlisting}[mathescape=true]
ChainVisible(L, fields, ca_cert, $\idlog$, #time, 
             sth_time) <=> 
 (Ex chain different_time #t.
  LoggerComputesAddLog(L, $\idlog$, fields, ca_cert, 
                       chain, different_time)@t 
  /* there has been an addition to $\color{sqBrown}\idlog$... */
  & #t < #time /* ...prior to #time and... */
  & not(sth_time = different_time) 
  /* ...that addition was visible */
 )

NoChainVisible(L, fields, $\idlog$, #time, sth_time) <=>
(not(Ex different_time #t. 
    LoggerComputesAddLogFields(L, $\idlog$, fields,    
                               different_time)@t
    & not(sth_time = different_time) 
    // there has never been a visible addition to the log
    & #t < #time
    ))
 // or: for all additions, the chain is invalid
 | ((All ca_cert different_time chain #t2 #t.
     LoggerComputesAddLog(L, $\idlog$, fields, ca_cert, 
                          chain, different_time)@t
     & not(sth_time = different_time)
     & #t < #time
     & ChainCheck(chain)@t2
      ==> ChainInvalid(chain)@t2
    ) 
    & (All different_time ca_cert chain #t.
    LoggerComputesAddLog(L, $\idlog$, fields, ca_cert, 
                         chain, different_time)@t
    & #t < #time
    ==> (Ex #t2. ChainCheck(chain)@t2)
      )
   )
\end{lstlisting}

These last two predicates might appear odd at first sight. When checking a certificate,
we will make use of them to distinguish whether a monitor can find a valid chain for a given pre-certificate entry in the log or not. By construction and if a monitor additionally verifies a found chain itself, a monitor is guaranteed to find a rogue certificate with a valid chain, if it exists in the log. This makes our model as expressive as a real-world CT monitor. For \lstinline|NoChainVisible|, we add an additional disjunction that captures the case where all chains for a given pre-certificate are invalid. Note that while invalid entries could be used to explore potential misbehavior too, we are interested in the accountability perspective. If the chain is invalid, we are not able to precisely blame the responsible CA. Invalid chains on the log are considered a log violation. This vector has similar effects as equivocation: log misbehavior leads to cover up of CA misbehavior.

\subsection{Fetching proofs and snapshots}
\label{sec:app_fetching_proofs_and_snapshots}

We keep inclusion and append-only proofs fully abstract as trace property. This means we do not compute proofs that are then verified when other parties receive them, but use fresh variables as symbolic proofs.
In order to be able to make use of proofs in the model, we use different trace properties that capture a snapshot of a log at the time of the proof.
We say that the proof succeeds in the model if an actual proof would have worked for the situation of the log at the time of creating the proof. We axiomatically assert this relation with restrictions on the trace.

For an inclusion proof of entry $x$, this means that the proof succeeds in the model if, at the time of handing out the proof for some log, $x$ is visible. Append-only proofs verify that between different snapshots of a log, no entries disappear. In \fvref{sec:inclusion_proof} and \fvref{sec:append_only_proof}, we present the details of necessary restrictions to model this approach specifically for inclusion and append-only proofs.
This approach assumes, as is common for the symbolic model, perfect cryptography of all Merkle tree proofs.

Fetching new snapshots consists of three steps. First, the party that fetches the snapshot starts by creating a |ProofRequest| fact. Loggers can receive this fact and then hand out the current snapshot for their log, if they are honest, or, if they are corrupted, for one of the logs that they maintain, including forged ones.

\begin{lstlisting}
rule ProofFetching_M:
 [ Fr(~session), !Monitor($M), !Logger($L, pkL) ] 
 --[ 
  FetchLogSnapshot($M, $L, ~session),
  FetchLogSnapshotTime($M, $L, ~session, $now),
  Time($now)
 ]->
 [ TLS_M_to_L($M, $L, ~session, $now)
   /* Note: The established connection is not persistent 
      such that new requests will be assigned a different 
      session identifier */
 ]
\end{lstlisting}

We add a |~session| value to every proof request in order to assign every start of a proof to the proof itself. In doing so, we are able to formulate time restrictions on a proof run, e.g., by stating that a client initiates the proof request after it has received a corresponding SCT.

In the second step, a logger hands out the snapshot for a $\idlog$ of their choice. This is denoted by the |LoggerHandsOutSTH| action where $\idlog$ is free.
$\idlog$ is dynamically instantiated and is therefore potentially attacker-controlled. We restrict that the instantiated $\idlog$ corresponds to a log that the logger maintains, i.e., the id exists. As honest loggers only maintain a single log, an attacker can only interfere here if the logger is corrupted.
The fact contains the session of the TLS connection allowing us to later reconstruct the temporal sequence of actions, involved parties, and a freshly generated |~sth| as well as the current time.

The current STH can be seen as a commitment of the log with $\idlog$. All underlying \lstinline|$log_entries| required for verifying the STH are delivered separately. 
\begin{lstlisting}[mathescape=true]
rule LoggerSnapshot_M:
 [ !Logger_s($\texttt{\$}$L, skL, pk(skL)), Fr(~sth) 
   /* abstract representation of the STH */, 
   TLS_M_to_L($\texttt{\$}$M, $\texttt{\$}$L, session, $\texttt{\$}$time) ]
 --[ 
  LoggerPresentsView($\texttt{\$}$L, $\texttt{\$}$M, $\idlog$, $\texttt{\$}$now, ~sth, session),
  LoggerHandsOutSTH($\texttt{\$}$L, ~sth, session),
  Time($\texttt{\$}$now)
 ]->
 [ !TLS_L_to_M($\texttt{\$}$L, $\texttt{\$}$M, $\texttt{\$}$snapshot, ~sth, session, $\texttt{\$}$time) ]
\end{lstlisting}

This rule is reused for all proofs involving the Merkle tree and monitors fetching all entries. This is due to our abstraction of proofs that we assume to be perfectly secure, which is typical for analyses in the DY model. We implicitly assume that loggers comply when presenting an STH and every underlying entry, as we cover malicious behavior separately. This assumption is relaxed when we turn to gossiping, where only the STH is shared, but not all underlying entries (\Cref{sec:gossiping}).

Incoming proof requests and outgoing proofs are not implemented on the network to exclude the Dolev-Yao adversary. Assume the contrary and an adversary that can intercept old proofs and replay them later, while the logger behaves honestly. In these cases, an old snapshot might be used at a later point and disprove inclusion. A natural step would be to blame the logger in this case, as the presented proof is not valid, but the logger was honest the whole time. This is problematic when we aim at achieving accountability. 

To highlight this limitation, we explicitly use a secure channel for these steps. Following the CT standard, the communication with a logger is implemented using HTTPS GET and POST requests, thereby excluding such attacks too. This secure channel has a session identifier that is used to relate the steps of the proof run.

In the third step, the handed-out snapshot is received and checked for the property of interest. We use different rules and restrictions representing every possible outcome of that check. We outline these in the next sections.%

\subsection{Inclusion Proof}\label{sec:inclusion_proof}

We formulate inclusion proofs using the visibility predicates presented in \Cref{lst:visibility_predicates-full}. The timepoint at which the log hands out the snapshot used in the proof, is then applied to the visibility predicate.

\begin{lstlisting}[mathescape=true]
restriction inclusionProofFails:
 "All IdFetcher L fields ca_cert sth session #t1. 
 InclusionProofFails(IdFetcher, L, fields, ca_cert, sth,
                     session)@t1
 ==> (Ex $\idlog$ time #t0. 
      LoggerPresentsView(L, IdFetcher, $\idlog$, time, sth, 
                         session)@t0 
      // Logger handed out an STH to the recipient...
      & #t0 < #t1 // ...prior to the check
      & EntryMissing(L, fields, $\idlog$, #t0, time))"
      // ...and the entry is not visible in that view
\end{lstlisting}

In the successful variant the entry is instead visible.

\begin{lstlisting}[mathescape=true]
restriction inclusionProofWorks:
"All [...] . 
 InclusionProofSucceeds($\id_\mathit{fetcher}$,L,fields,ca_cert,
                        chain, sth, session)@t5
 ==> Ex id $\idlog$ time #t2.
  LoggerPresentsView(L,id,$\idlog$,time,sth,session)@t2 
  & #t2 < #t5
  & EntryVisible(L,fields,ca_cert,chain,$\idlog$,#t2,time)"
\end{lstlisting}

Depending on the action fact in the trace, we learn whether the proof can succeed.

\subsection{Append-Only Proof}\label{sec:append_only_proof}

When we talk about the append-only property, we always refer to a pair of snapshots of a log. The proof shows that all entries in the older snapshot are also included in the more recent variant. 
This proof should not be possible if entries have been removed from the log in between.

We capture this behavior by introducing a rule that simply collects and stores a snapshot of the log and a second kind of rule that either proves or disproves the property based on all held snapshots. Similar to the inclusion proof, we use restrictions for that.

\begin{lstlisting}
rule MonitorStoresSnapshot:
[ !TLS_L_to_M($L, $M, $snapshot, sth, session, $time),
  !Logger($L, pkL), !Monitor($M) ]
--[ 
 SnapshotStored($M, $L, session, sth),
 Time($now)
]->
[  ]
\end{lstlisting}

The implementation for a client doing the same is analogous.
The formalization of an unsuccessful append-only proof is given below. Note that for a violation it is sufficient to find one pair that witnesses the violation.

\begin{lstlisting}[mathescape=true]
restriction AppendOnlyFails:
"All sth session id L #t6. 
AppendOnlyViolation(id, L, sth, session)@t6
/* Append-only violation found by id with violating L is 
   only possible if a counterexample is found */
==> 
 (Ex fields ca_cert chain time t ${\idlog}_1$ ${\idlog}_2$ time2 sth2 
     session2 #t0 #t1 #t2 #t3 #t4 #t5. 
  LoggerComputesAddLog(L, ${\idlog}_1$, fields, ca_cert, 
                       chain, t)@t0
  & EntryVisible(L, fields, ca_cert, chain, ${\idlog}_1$, 
                 #t2, time) 
  /* there is an entry in the log that is visible at 
     #t2 and time period time in ${\idlog}_1$ */
  & FetchLogSnapshot(id, L, session2)@t1 
  /* someone starts fetching a log view */
  & LoggerPresentsView(L, id, ${\idlog}_1$, time, sth2, 
                       session2)@t2 
  /* The Logger presents the view for $\color{sqBrown}{\idlog}_1$ */
  & SnapshotStored(id, L, session2, sth2)@t3 
  /* this view is received and stored by id */
  & FetchLogSnapshot(id, L, session)@t4
  /* a second view is fetched by id (later) */
  & LoggerPresentsView(L, id, ${\idlog}_2$, time2, sth, 
                       session)@t5
  /* the logger possibly presents a different log */
  & EntryMissing(L, fields, ${\idlog}_2$, #t5, time2)
  /* one entry that was visible in the past is now not 
    visible anymore -> clear append-only violation */
  & #t0 < #t1 & #t1 < #t2 & #t2 < #t3 & #t2 < #t4 
  & #t4 < #t5 & #t3 < #t5 & #t5 < #t6)"
\end{lstlisting} 

For verifying the append-only property given an up-to-date snapshot of the log, we take all existing snapshots a party has retrieved and stored in the past into account. The formalization is analogous to the previous one by asserting that every entry that was visible before should still be visible.

    \section{CT with Receipts (Details)}
\label{sec:ct_receipt}

So far, we have excluded monitor misbehavior from our accountability analysis. 
CT does not provide us with a mechanism suited for accountability tests, as a party distinct from the monitor cannot verify whether the monitor purposefully ignored a rogue certificate, let alone whether a monitor even saw the certificate in the first place.

To address this limitation, we further extend our model of SCT Auditing with \emph{receipts} issued by the monitor back to the logger. The receipt proves to a logger (and any other party in possession of it) that the monitor has seen a specific snapshot of the log.
SCT Auditing is in particular of interest here, as we were able to show that accountable authenticity can be achieved in SCT Auditing given honest monitors, while allowing corrupted loggers. With this receipt approach we aim to achieve accountable authenticity even in the presence of corrupted monitors.

We adapt our model by giving a key pair to every monitor and model the signing process.

\begin{lstlisting}
rule signReceipt:
let
 receipt = sign(<$M, chain>, skM)
in
[ CraftReceipt($M, chain), 
  !MonitorKey_s($M, pkSKM, skM) 
]
--[ Time($now) ]->
[ !TLS_M_to_C(receipt) 
  /* We already assumed a TLS connection 
     between client and monitor before */ ]
\end{lstlisting}

\lstinline|CraftReceipt| is a fact emitted by the SCT Auditing rules when a monitor receives an audited certificate chain from a client. The receipt is a simple signature over that received chain and the monitor's identifier. The signature is then sent back to the client via a TLS channel, modeled by the \lstinline|TLS_M_to_C| fact.

The client then, if necessary, is able to forward the receipt to the domain owner. 

\begin{lstlisting}
rule domainOwnerBlamesMonitor:
let
 ca_fields = 
    <$rootCA, $CA, pkCA, 'pubKey-CA', snCA, $i0, $j0>
 pub_fields = 
    <$CA, $id, pkID, 'pubKey', sn, $i, $j, sct1, sct2>
 fields = <$CA, $id, pkID, 'pubKey', sn, $i, $j>
 chain = <<pub_fields, pub_sig>, <ca_fields, ca_sig>>
in
[ In(<$M, receipt, chain>), !MonitorKey($M, pkSKM), 
  !TrackSubject_Contract($M, $idServer), 
  !Root_CA($rootCA, pkRootCA) ]
--[
  Eq(verify(receipt, <$M, chain>, pkSKM), true()), 
  Eq(verify(pub_sig, pub_fields, pkCA), true()),
  Eq(verify(ca_sig, ca_fields, pkRootCA), true()), 
  DO_MonitorBlame($M, $CA, ca_fields, fields),
  Asserts($id,'This monitor did not react', $M,<fields>),
  Asserts($id, 'That intermediary CA signed the following incorrect statement and the validation chain says the following about that CA.', $CA, <fields, ca_fields>),
  // external validator style assertions for authenticity
  RogueCert($CA, fields),
  Time($now)
]->
[ ReceiptChecked($M, fields, ca_fields) ]
\end{lstlisting}

In general, we assumed in SCT Auditing that every received certificate is checked.
To allow corrupted monitors to diverge from this behavior, we allow them to \emph{sleep} on an entry, i.e., seeing the entry but not acting on it. In this case, the \lstinline|DO_MonitorBlame| action can be reached.

Further rules, omitted here, consider the variants where we cannot blame a monitor. This is the case if either the monitor acted benignly when confronted with the rogue certificate, i.e., raised an alert to the domain owner, or if the monitor did not need to react, as the certificate was benign.

\subsection{Accountability Analysis}
\label{sec:app_ct_receipt_acc_analysis}

In this scenario, we assume that a domain owner finds a rogue certificate that was actively used. As the domain owner has access to the ground truth, accountable authenticity via an external check is easily achieved, as seen in \Cref{sec:acc_auth_pki}. But, what if the domain owner contracted a monitor service to do the same thing, but it raised no alarm?

The CT-receipt model allows tests that detect monitor misbehavior and, combined with previous test components concerning the intermediate or root CA misbehavior, provides us with accountable authenticity against untrusted monitors. This allows, e.g., for probabilistic checking of the monitor, or to use multiple monitors to validate each other (which is essentially the same test).

\begin{lstlisting}
test receipt_1:
 "Ex M id fields rootCA ca_fields #t #t1 #t2 #t3.
 SCTReceiptCheck(M, fields)@t
 & Asserts(id, 'That intermediary CA signed the following incorrect statement and the validation chain says the following about that CA.', CA, <fields, ca_fields>)@t1 
   // CA blamed
 & Asserts(id, 'This monitor did react', M, <fields>)@t2 
   // monitor not blamed
 & Asserts(rootCA, 'I signed this intermediate CA.', CA, 
          <ca_fields>)@t3" 
   // CA is liable
\end{lstlisting}
\begin{lstlisting}
test receipt_2:
 "Ex M id fields CA ca_fields #t #t1 #t2 #t3.
 SCTReceiptCheck(M, fields)@t
 & Asserts(id, 'That intermediary CA signed the following incorrect statement and the validation chain says the following about that CA.', CA, <fields, ca_fields>)@t1 
   // CA blamed
 & Asserts(id, 'This monitor did react', M, <fields>)@t2 
   // monitor not blamed
 & Asserts('CCADB', 'No root CA signed this CA.', CA, 
           <ca_fields, rootCA>)@t3" 
  // rootCA is liable
\end{lstlisting}

Two more analogous tests blame the intermediate CA instead, as seen before.


Together they construct the following lemma, which verifies.

\begin{lstlisting}[caption = {SCT Auditing with receipts provides acc.\ auth.},
label=lst:accountable-authenticity-sct-receipt]
lemma A_Receipt:
receipt_1, receipt_2, receipt_3, receipt_4
accounts for
 "All CA fields rootCA ca_fields L1 L2 M #t0 #t1.
 SCTReceiptCheck(M, fields)@t0
 & BrowserAcceptsChain(CA, fields, rootCA, ca_fields, 
                       L1, L2)@t1
    ==> (Ex CA #t. GroundTruth(CA, fields)@t)"
\end{lstlisting}

\subsection{Discussion}

Using receipts issued by monitors to loggers, we were able to design tests that allow us to blame corrupted monitors that purposefully ignored rogue certificates. Together with previous tests blaming corrupted CAs, we achieve accountable authenticity even in the presence of corrupted monitors. Receipts illustrate a possible mechanism for holding monitors accountable for their actions. 
We moved multiple tests to the domain owner here, which is a natural choice as the domain owner has access to the ground truth and is particularly interested in correct behavior of the contracted monitor.

We relied on the fact that the domain owner can access such a receipt, e.g., because the client forwards it. It is not straightforward how to ensure that the domain owner actually receives such receipts in practice. While domain owners run their own web server, i.e., they are addressable, sending the receipt to the \emph{correct} web server boils down to the authenticity problem we started with. A more realistic perspective is that domain owners purposefully act as clients, to probabilistically test for misbehavior. This would still need `normal' clients to require receipts, but they would not need to always forward them.

Additionally, we assumed that the monitor issues such receipts for every received certificate. While a corrupted monitor could choose not to issue a receipt, note that a client could always demand a receipt or otherwise refuse to accept the certificate to protect themselves.

As complexity on the domain-owner side increases, one could raise the argument that a monitor distinct from the domain owner is not necessary anymore. Recall that everyone can become a monitor and self-run monitors have the advantage that the running party immediately trusts this monitor. 
However, SCT Auditing assumes the presence of a powerful monitor that has access to all certificates that clients eventually see. We already discussed in \Cref{sec:sct_auditing} that self-run monitors with these capabilities are not realistic. A mechanism to hold monitors accountable is in particular interesting if domain owners have to rely on such a powerful third-party monitor.
 
    \section{Modeling STH
Gossiping (Details)}\label{sec:app-modelling-ct-gossiping}

In this section, we will introduce the additions to our model and reformulate properties to apply them to STH gossiping.

If a monitor wants to ensure that it can see the entries behind a gossiped STH, it additionally needs to run an append-only proof between the two snapshots it considers.

We use similar restrictions to axiomatically assert append-only proofs (\fvref{sec:append_only_proof}), but this time the gossiped proof is not necessarily handed out to the monitor but rather to any party to reflect the nature of gossiping.

\begin{lstlisting}
rule monitorFindsAppendOnlyViolation_Gossiped:
[ !LoggerDict($M, $L), 
  /* Logger is monitored */
  !TLS_C_to_M($C, $M, $gos_snapshot, gos_sth, gos_sess), 
  /* Gossiped snapshot submitted by a client */
  !TLS_L_to_M($L, $M, $snapshot, sth, sess, $time)
  /* Own snapshot of the logger that the monitor holds.
     We assume that this snapshot is freshly fetched 
     (see  restriction below), s.t. the monitor can 
     verify that it can see as much as the client. */
  ]
--[
 Gos_AppendOnlyViol($M,$L,sth,sess,gos_sth,gos_sess), 
 /* see restriction */
 Asserts($M, 'The log has violated the append-only property between these two snapshots.', $L, 
         <gos_sess, sess>),
 MonitorReceivesGossip($M, $L, gos_sess, sess),
 Time($now)
]->
[ AppendOnlyChecked($M, gos_sess, sess) ]
\end{lstlisting}

The corresponding restriction relaxes the assumption that both snapshots have been handed out to the monitor to reflect the idea of gossiping.

\begin{lstlisting}[mathescape=true]
restriction appendOnlyFails_Gossiped:
"All M L session sth gossiped_sth gossiped_session #t5. 
 Gos_AppendOnlyViol(M, L, sth, session, gossiped_sth, 
                    gossiped_session)@t5
 ==> 
  (Ex id1 fields ca_cert chain time t time2 $\idlog$ ${\idlog}_2$ 
   #t0 #t1 #t2 #t3 #t4.
   LoggerComputesAddLog(L, $\idlog$, fields, ca_cert, 
                        chain, t)@t0
   & EntryVisible(L,fields,ca_cert,chain,$\idlog$,#t2,time) 
   // There is an entry that was visible...
   & FetchLogSnapshot(id1, L, gossiped_session)@t1
   & LoggerPresentsView(L, id1, $\idlog$, time, gossiped_sth, 
                        gossiped_session)@t2 
   // ... for a given proof at #t2
   & FetchLogSnapshot(M, L, session)@t3
   & LoggerPresentsView(L,M,${\idlog}_2$,time2,sth,session)@t4 
   /* second view is handed out to the monitor directly 
      (only one of both views is gossiped as assumed) */
   & EntryMissing(L, fields, ${\idlog}_2$, #t4, time2)
   /* one entry that was visible in the gossiped view is 
      now not visible anymore -> clearly a violation */
   & #t0 < #t1 & #t1 < #t2 & #t2 < #t3 
   & #t3 < #t4 & #t4 < #t5
)"
\end{lstlisting}

The other rules consider the opposite case and are analogously adapted.

\end{full}
\end{document}